# Right and wrong: ten choices in language design

## Bertrand Meyer

**Constructor Institute and Eiffel Software**

Bertrand.Meyer@inf.ethz.ch

## Abstract

A description of language design choices that have a profound effect on software quality, criticism of how ordinary OO languages addressed them, and explanation of the thinking behind Eiffel's corresponding mechanisms.

## A Introduction

Object-oriented programming rules the world. Or so it seems.

In reality, while languages presented as object-oriented (OO) have come to enjoy widespread usage, their usage frequently shows OO methodology being misunderstood and misapplied. Some of the blame lies with the design of the languages themselves.

Eiffel as a programming language is an attempt to help programmers apply the methodology of OO analysis, design and programming, including Design by Contract (DbC) as expounded in [4] and [11]. Eiffel's design is a combination of object-oriented concepts with ideas from formal methods (particularly axiomatic semantics), correctness-by-construction, modularity, and more generally modern software engineering. At the core of this approach are a set of explicitly stated principles, from Design by Contract to the Open-Closed and Single Choice principles and a good dozen others, which the supporting methodological publications explain in detail.

Other OO languages are not so firmly rooted in methodology. C++, in particular, had compatibility with C as one of its key design goals and presents itself as a "multi-paradigm language" (its creator also described it as a "Swiss knife") rather than strictly an OO language. Its successors retain a C-like style and a design guided by other criteria.

Half-embracing the object-oriented paradigm has negative consequences on software quality: OO design is a consistent discipline which needs to be applied consistently to yield advantages of extendibility, reusability, scalability and reliability. The programming language can support this goal, or not. This article identifies Eiffel's resolution of ten individual language issues, and compares some of these choices with those of other widely used OO languages, showing how their misunderstanding of methodological principles can lead programmers to produce bad designs and programs of subpar quality.

## A.1 Background

The need for this article arose in part from the new difficulty of teaching programming.

It used to be that few people in society knew anything about programming. Among those who did have programming experience, even fewer knew anything about OO programming when it started attracting attention. At that time, as one now realizes, teaching the concepts was fairly easy: they are consistent, powerful and convincing. No preconception on the learners' side stood in the way of accepting them with an open mind.

Today, having some programming experience is no longer a rare skill. Increasingly, pre-university curricula include some programming. Many people have even encountered an OO language; they include most students starting a university computer science program. Language exposure is not by itself the problem; the trouble is that together with the language they were taught bogus principles such as "*multiple inheritance is bad*" (often just "*inheritance is bad*"), or "*overloading is an OO mechanism*" or "*you should make fields private and write getter functions to provide access to them*", or "*you can return from anywhere in a function*", or "*writing interfaces helps OO design*" or "*you should use defensive programming*" and other atrocities apparently entered into young minds at the same time and with the same authority as the information that the earth revolves around the sun.

Teaching is hard enough, but the first task when dealing with such preconceptions is to un-teach, which is even harder. This article is an attempt to set basic choices straight.

## A.2 Design guidelines

One of the goals of the following discussion is to highlight the mode of thinking that has guided the design of Eiffel.

There seems to be a general trend in modern language design of proposing clever constructs that offer impressive modes of expression for certain cases. While this quest for programmer convenience is legitimate, it should only be one of the guiding forces in language design. The criterion that plays the principal role in the design and evolution of Eiffel is the constant effort to help programmers produce software that meets fundamental software quality criteria [4] [11]: correctness, robustness, extendibility, reusability.

Not *preventing* programmers from writing *bad* programs (that is not possible — at best the programming language may be able to make bad programs harder to produce) but *helping* them write *good* programs if they are determined to do so.

This observation also serves to qualify the focus on quality: Eiffel is in no way intended as a straitjacket, a holier-than-thou injunction to programmers to go through endless hurdles in pursuit of correctness. To the contrary, the language should be supportive of the programmer, devoid of verbosity, and pleasant to use, which is where the expressiveness criterion comes back. Sections 1.5 and 7 will explore this aspect, showing how Eiffel reconciles advanced OO concepts with concise traditional notations.

## A.3 Discussion style

The presentation follows the tradition of articles of a generation ago, discussing concepts of programming methodology and programming languages.

It does not resemble most articles that get published today, and would be desk-rejected by all the significant venues because of its absence of an empirical evaluation. While quantitative assessments are of great value, there remains room for discussions of a conceptual nature, based on logical arguments rather than experiment results. (At the very least, they provide experimenters with something to experiment on.)

## A.4 Language criticism

The strong terms in which the discussion assesses some design decisions of dominant languages do not imply disrespect for either the designs or the designers. The languages' success is evidence enough that they filled a need. But it does not forestall criticism.

Discussions of such matters are healthy. They are also important: at issue are not opinions and susceptibility, but the quality of software driving today's societal processes. A harmful language design decision leads to buggy programs and to malfunctions that ultimately affect people in ways sometimes superficial but sometimes existential.

## A.5 Scope

After introducing a few mathematical conventions in section B, the discussion covers the following ten design choices:

- Uniform access, or how to use object properties abstractly (section 1).
- Control structures of structured programming (section 2).
- Overloading and constructors (section 3).
- Inheritance, single and multiple, and the noxious notion of interface (section 4).
- Design by Contract mechanisms (section 5).
- How to handle static and shared information (section 6).
- How to provide classic syntax for novel concepts (section 7).
- The proper role of exception handling (section 8).
- Concurrent programming in an OO context: SCOOP (section 9).
- Removing null pointer dereferencing (section 10).

In addition, Appendix A1 has some observations about language definition and Appendix A2 about how to manage the *evolution* of a language.

### A.6 The role of the type system

Eiffel's type system governs most of the mechanisms covered here, from creation and conversions to concurrency and void safety. Together with Design by Contract concepts, it is the essence of Eiffel, one might almost say its soul. The first publication about Eiffel [2] described the subtle combination of genericity and (multiple) inheritance that forms its backbone. All major subsequent evolutions are extensions and tightenings of the type system, devised to achieve a proper balance of expressive power and safety. Programming in Eiffel largely consists of putting that type system to work.

### A.7 Precise object-oriented terminology

A *class* (static mechanism) defines a set of potential run-time *objects* (dynamic concept), its *instances*. A class has *features* (or "members") specifying operations applicable to its instances. A feature may be a *command*, changing objects, or a *query*, providing information about objects. A command is also known as a *procedure*. A query may be implemented as a *function*, computing the required information, or an *attribute*, which defines a *field* present in every object and containing that information. (In the Java/C# context the term "field" also covers attributes, but it is preferable to distinguish the static and dynamic views: a class has attributes, defining fields in its instances.) Procedures and functions are both defined by algorithms and are known as *routines* (or "methods").

A program element can be correct at three levels (corresponding to syntax, static semantics and semantics), each only defined if the preceding one is: it is *syntactically correct* if it is built according to the structure of the language (for example, matching parentheses); *valid* if it satisfies additional static rules not expressible in syntax formalisms (for example, type rules); and *correct* if its execution produces the expected results.

## B MATHEMATICAL NOTATIONS

While largely informal, the discussion will at times reflect the underlying mathematical concepts, particularly for the discussion of inheritance in section 4.4, using the following conventions (see [24] for the outline of a general mathematical basis for programming).

### B.1 Functions

A function is a set of pairs, for example $\{[Elizabeth, Philip], [Charles, Diana], [Ann, Mark]\}$ (since it is a set, not a sequence, the order in which we list these pairs is immaterial), such that for any given element at most one pair has it as its first element. Removing this restriction, for example by including a pair $[Charles, Camilla]$, still yields a *relation*, but it is no longer a function.

The sets from which a function's pairs take their first and second values (respectively) are its *source set* and *target set*. The actual first and second elements of pairs in the function make up its *domain* and *range* (respective subsets of the source and target sets); if we call the preceding function $spouse$, its domain is $\{Elizabeth, Charles, Ann\}$ and its range $\{Philip, Diana, Mark\}$. These concepts also apply to relations that are not functions.

The defining property of functions makes *functional notation* possible: if $x$ is an element of the domain of *spouse*, we may use *spouse* $(x)$ to denote the single element $y$ such that there is a pair $[x, y]$. For example *spouse* (*Elizabeth*) is *Philip*.

We are particularly interested in functions with a *finite* domain, and hence finite range; the set of such functions with source and target sets $A$ and $B$ will be written $A \nrightarrow B$. "*Mapping*" is a synonym for "function" but will be restricted in this presentation to finite functions only.

If $f$ and $g$ are functions, their composition $f\,;\,g$ is the function $h$ such that $h\,(x) = g\,(f\,(x))$ wherever defined. (It is also commonly written $g \circ f$, but the ";" notation is convenient since it lists operands in order of their application.)

## B.2 Overriding union

An important mathematical operator for the modeling of programming concepts is "overriding union" $\Cup$, which yields a function from two functions.

The union of two functions is a relation but not necessarily a function; for example if *spouse_new* is {[*Edward*, *Sophie*], [*Charles*, *Camilla*]}, then *spouse* ∪ *spouse_new* has two pairs starting with *Charles*, making it violate the defining property of functions.

To guarantee that the result is a function, we may instead of union use the $\Cup$ operator. It acts like ∪, but in case of a pair clash it only retains the pair from the second operand, so that it always yields a function. (The notation reflects this property by including a "\" which "leans" towards the second operand.) Here *spouse* $\Cup$ *spouse_new* is the function {[*Elizabeth*, *Philip*], [*Charles*, *Camilla*], [*Ann*, *Mark*], [*Edward*, *Sophie*]}. Overriding union is useful to model situations, such as inheritance, in which a new mapping both complements and overrides an existing one.

## B.3 Finite permutations

We will need the notion of finite permutation on a set $A$. It is the usual notion of permutation (a total function $\sigma$ from $A$ to $A$ that is one-to-one, meaning that both $\sigma$ and its inverse $\sigma^{-1}$ are injective functions of domain $A$), but with the condition that it is the identity except on a finite subset of $A$. An example is $\{[a, b], [b, a], [c, d]\} \cup \{[x, x] \mid x \notin \{a, b, c\}\}$; in other words, a function that maps $a$ to $b$, $b$ to $a$, $c$ to $d$, and any value other than $a, b, c$ (including $d$) to itself. Another way of stating that convention is that we only consider permutations that substitute a finite subset of elements, extending them with the identity on the rest of $A$ to so as to get total functions.

# 1 UNIFORM ACCESS

We start the review of programming language choices with a matter on which the original OO language, Simula 67, had things right and the design of C++, faithfully followed by today's dominant languages, introduced a conceptual mistake breaking the consistency and simplicity of OO principles. Even UML, which as a modeling language should rely on abstract concepts rather than implementation nuances, perpetuates the flaw. In some cases, such as C#, the designers realized the awkwardness of the result but instead of adopting the obvious simple solution they devised a contorted workaround.

## 1.1 Computing, or looking up

Uniform access, the elegant principle violated by these designs, is the idea [4] [11] that "*it does not matter whether you look up or compute*". Client modules needing a certain element of information should be able to obtain it in the same way regardless of the implementation choice: *storage* (the information is recorded in the corresponding data structure, obtaining it is a simple lookup); or *computation* (the information can be computed through an algorithm, obtaining it requires executing the corresponding operation).

Uniform access is a natural consequence of the general principle of data abstraction, at the basis of object technology. The principle (also known as encapsulation, and closely related to "information hiding") states that clients of a module should be able to access its elements of functionality through the specification of their abstract properties, without having to rely on properties of their implementation. These functionality elements are features (operations), known to client programmers through their specifications only. Features are of two kinds: commands, which can change the state of objects; and queries, which return information about objects. Uniform access governs queries.

A typical query, for a bank account class, is the operation *balance* returning an account's balance. Consider an implementation in which an account object looks like this ("Representation A"):

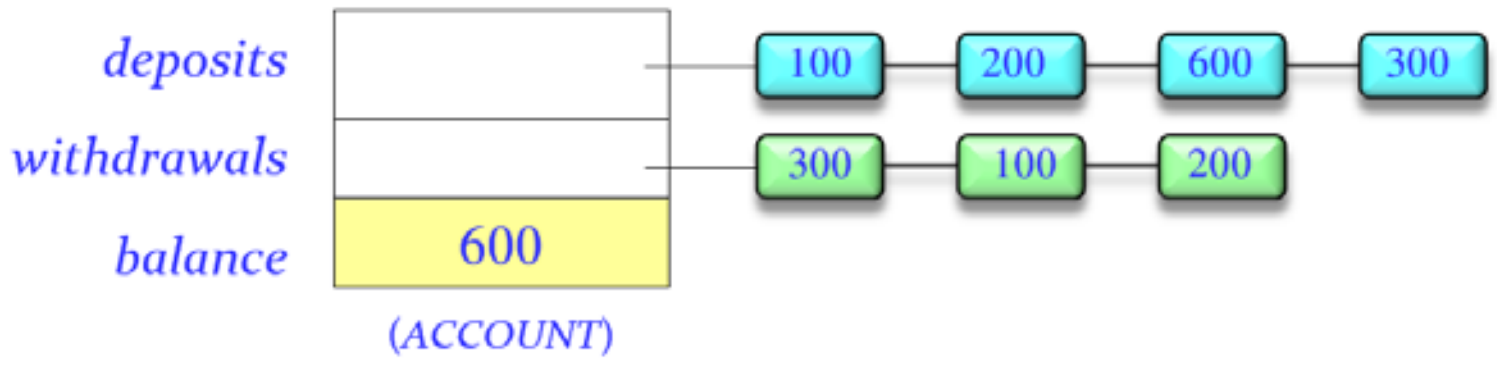

For every account object, the representation includes (in addition to other properties not shown such as account number and owner), a field denoting the current balance and two lists of banking operations, one for deposits and the other for withdrawals. Each element of these lists includes the amount involved (plus any other information not shown in the figure, such as the operation's date). Representation A has the following properties:

- The fields are not independent. They are constrained by a property which we must include in the class invariant: *balance* = *deposits*.*total* – *withdrawals*.*total* (assuming a function *total* yielding the accumulated value of such lists of banking operations).
- Commands *deposit* and *withdraw* must maintain this property by updating both the *balance* field and the respective list. For example, *withdraw* (100) must not only decrease the *balance* field by 100 but also add an element of value 100 to the list *withdrawals*.
- To obtain the balance, it suffices to look up the *balance* field, which is always up to date thanks to the invariant property.

In another approach, which we may call "Representation B", the representation of an account does not include the *balance* field. Representation B can still provide all the functionality of the original but internally does things in a different way:

- Commands *deposit* and *withdraw* only need to add an element to the respective list. There is no field to update.
- Finding out the balance of an account, on the other hand, now requires a computation, returning the value of *deposits*.*total* – *withdrawals*.*total*. Computing *total* involves (in a simple implementation) traversing the respective list to sum its element values.

Uniform access means protecting client modules from such internal implementation decisions as the choice between Representation A and Representation B. Languages such as C++, Java and C# force a different syntax to access the balance of an account *a*:

- *a*.*balance* for a field representation (A).
- *a*.*balance* () for an representation by a function (B).

While the difference is small, it forces knowledge of suppliers' implementation onto client programmers and negates the benefits of information hiding. Changes of representation between variants such as A and B occur routinely during the development of a system; preserving clients against having to update each time is part of the general discipline of data abstraction and key to the stability of the development.

It gets worse.

## 1.2 What do we do with fields?

Even the most uneducated users of languages such as Java, C++ and C# have heard about information hiding and realize that giving clients unfettered access to fields of objects (attributes of the corresponding classes) is not good for their karma. Hence the advice universally taught in the manuals for such languages (which its recipients mistake for good OO methodology, rather than for what it is — punishing programmers with the consequences of flawed language design decisions): *never export an attribute*. Instead, programmers are enjoined always to write a "getter function" of the form

```
balance_getter: INTEGER                                    -- /P1/
        -- Balance of this account.
    do
        Result := balance
    end
```

and export only *balance_getter*, keeping *balance* secret. (The example uses the syntax of Eiffel, where **Result** denotes the result to be returned by the function; see below.)

Why this self-imposed rule? The reason is a property that is more crazy than any seen so far (although yet crazier ones follow): if you export an attribute in C++ and its successors, you export it with *full rights for the client*. So if *balance* were exported, any client code could use, for an account *a*:

*a*.*balance* := 5000 -- /P2/

which causes a gross violation of information hiding since it allows clients to play at will with the internals of a supplier object, in a way similar to what happens if instead of using an electronic device through its buttons you open it up and play with the wires. (There is a reason the device's notice says that doing so voids the warranty.)

The example is typical of why such unrestricted exporting is wrong: the designer of class *ACCOUNT* has clearly decided, regardless of the implementation, that clients should not directly modify *balance* but go through the commands *deposit* and *withdraw*, which may perform various checks and updates. In Eiffel, they will have preconditions, including that the argument should be non-negative (**require** *amount* ≥ 0) and, for *withdraw*, not too large (**require** *amount* ≤ *balance* – *authorized_overdraft*).

With remote assignments to *a*.*balance* allowed, all such methodological protections crumble.No wonder then that exporting attributes is widely presented as a cardinal sin.

This systematic advice is, in itself, a bad sign: while it is normal to associate methodological guidelines with language constructs, a language design that includes a mechanism together with *an express injunction not to use it* can only described as sado-masochistic.

## 1.3 Exporting queries as queries

The proper and obvious approach is to allow the exporting of a query, whether an attribute or a function — but not *as an* attribute or *as a* function, which is none of the clients' business: as a query. In the interface specification of an Eiffel class, you will see

*balance*: *INTEGER* -- /P3/
    -- Current balance of this account.

whether the class internally uses Representation A or Representation B (or another). This specification is generated automatically from the class text, through environment tools that produce the "contract view" of a class (4.8 below). It is the same for an attribute or a function and does not betray which of these cases hold, but in both of them will include abstract properties from preconditions, postconditions and invariants, for example the invariant clause

*balance* ≥ *authorized_overdraft* -- /P4/

In both cases, a client class may use the notation *a*.*balance* — if *ACCOUNT* exports *balance* to it — and use it as an expression. One cannot assign to an expression ($a + b := 1$ makes no sense in usual programming languages), so the remote assignment *a*.*balance* := 5000 is trivially incorrect at the syntactic level, not even requiring a validity rule. (See 1.5 below for a legitimate re-interpretation of this notation.)

A query may have pre- and postconditions (for example **require** *is_open* for *balance*), which might seem to require getter functions, but do not: in line with the Uniform Access principle, Eiffel supports pre- and postconditions for attributes as well as functions.

## 1.4 Getters and setters

Once one has understood that it is OK to export a query — attribute or function — when exporting it *only* as a query, the notion of a getter function becomes useless. Yet code in C++ and its successors is awash with getter functions such as *balance_getter* above. Programmers writing such code think they are applying OO methodology but in reality they are only, as victims of confused language design, creating code bloat.

It is remarkable that recent programming languages have tended to terseness in their syntax, favoring braces and other special symbols, in the C tradition, over keywords — but then cause programmers to produce considerable amounts of such boilerplate code, doubly decreasing readability. Code bloat is always harmful, leading not only to waste of time in the initial coding effort but to an increased maintenance burden ("technical debt"). Concision is not served by cryptic syntax but by getting rid of useless code.

## 1.5 Assigner commands

Getter functions are always spurious and harmful. Setters are another matter: for an exported attribute, the class author needs to control what kind of change to permit clients to perform. In the *balance* case for a bank account, it is unlikely that the class author would permit clients to set the corresponding field of an *ACCOUNT* object to an arbitrary value; the appropriate policy is probably, as assumed above, to define two setter commands *deposit* and *withdraw*, which do modify *balance* but each in its own controlled way — add or subtract a non-negative value, within certain bounds. Note the importance of equipping such setter with contracts, for example the precondition in

```
withdraw (amount: INTEGER)                                    -- /P5/
    require
        amount ≤ 0
        amount ≥ balance – authorized_overdraft
    …
```

In this case the only way to modify the balance of an account is to call *withdraw* or *deposit*, both of which perform modifications of specifically permitted kinds.

In other cases the author of a class may want to allow clients to set the value directly; for example a class *HEATER* may have an attribute *temperature* and a setter command

```
set_temperature (t: REAL)                                     -- /P6/
        -- Change the temperature to t.
    require
        t ≥ 15
        t ≤ 25
    do
        temperature := t
    ensure
        temperature = t
    end
```

Even when a class lets its clients directly set the value of a field, it may restrict the range of possible values through contracts (particularly preconditions) as in this example.

In such cases Eiffel offers a notational facility preserving a classical, intuitive notation as a shortcut for a methodologically sound operation. If you add an "**assign**" specification to the declaration of *temperature*, making it

*temperature*: *REAL* **assign** *set_temperature*

you are associating *set_temperature* as an "assigner command" for *temperature*, with the effect that for the setting call

*my_heater*.*set_temperature* (21.5) -- /P7/

clients can also use a synonymous syntax:

*my_heater*.*temperature* := 21.5 -- /P8/

/P8/ is syntactically identical to /P2/ (a remote assignment, in C++ etc., to a field of an object) but is not an assignment. Known as an "assigner call", it is in fact a procedure call, like /P7/, but using a different syntax. It is bound by the same type rules (for the type of the argument in /P7/ and the right-side expression in /P8/) and, just as importantly, the contract: the precondition defined in the declaration of *set_temperature* continues to apply. (C# has a facility in the same spirit, "properties", but it does not remove the need for getter functions, and of course does not support any notion of contract.)

The use of classical syntax to represent constructs that fit in OO methodology is part of a systematic *language design* pattern, discussed further in section 7.

## 2 DIJKSTRA WAS RIGHT

Before continuing with object-oriented abstraction mechanisms we briefly stop with a plague that should have been eradicated by our grandparents but still affects us. It is a common myth that "*Python* (*like almost every programming language today*) *supports structured programming which controls flow using if/then/else, loop and subroutines*" (from a comment in a Stack Overflow discussion at bit.ly/3ErsDPZ). "Myth" is a polite way to say "lie": all major imperative programming languages today other than Eiffel have an essentially unrestrained goto, and programmers use it generously. It is not expressed as a "goto" but takes the form of:

- Instructions to break out of a loop: break and continue.
- Worse yet, the "return" instruction in routines.

(Plus exceptions, discussed in section 8.) It is hypocritical and misleading to pretend that these languages forsake the goto and apply "structured programming" since these three constructs correspond to all practical uses of the "goto" other than totally demented ones. They constitute a cynical violation of the principles enunciated in Edsger Dijkstra's famous 1968 goto diatribe, and specifically of the structured programming injunction to organize the control flow into one-entry, one-exit blocks (OEOE).

OEOE is particularly important in the case of routines. An instruction (in Python, Java, C++, C#…) of the form

**return** *exp*

may appear anywhere in the body of a routine, breaking out of all remaining computations so as to return (for a function) the value of *exp*. Such a jump operation is worse than the others since it is a goto reaching *across procedural boundaries*.

Eiffel applies a strict OEOE policy; there is no goto of any kind or disguise. Functions return their values in a simple fashion: they use the predefined local variable **Result**, initialized to the default value of the function's return type (0 for integers, false for booleans etc.). Whatever value **Result** has at the function's single exit point (the syntactical end of its text) is the value it returns . This convention avoids the need to check that a function's result is always defined (as required with goto-style "return" instructions).

Dijkstra's arguments were informal, largely an appeal to gut feeling ("*for a number of years I have been familiar with the observation that the quality of programmers is a decreasing function of the density of go to statements they produce*" — one can hear the referees screaming "*where is the empirical study?*"). Even the more respectable argument that structured-programming proponents made at the time — that **goto** complicated formal axiomatic reasoning — went away when someone pointed out that it suffices to perform, at each control-flow node, a logical **or** of the conditions at predecessor nodes. The present discussion also does not include a scientific demonstration but relies on the author's (and the Eiffel community's) decades of programming without a branching instruction. Not as a methodological rule (which anyone can violate, as programmers in other languages constantly do) but as a strict *impossibility* since no goto, actual or ersatz, is available.

The lesson, even if not scientifically established, is clear: there are only benefits in sticking to OEOE. Program flow becomes simpler and clearer. When you have something more complicated, and your Java programmer friends would tell you "oh, just break out of the loop", you know better. You either:

- Introduce a boolean local variable to express the exit or continuation condition, which makes the program reader clearly understand what is going on logically.
- Realize that the structure is getting too complicated — time to modularize.

These observations apply both to loops and to routines. In the case of routines, the guarantee that the code is executed to the end, and that there is no need to look for possible short-circuits or early exits, considerably simplifies building, reasoning and debugging.

## 3 OVERLOADING AND CONSTRUCTORS

Name overloading, which we may more precisely call *syntactic* overloading because it is a notational convenience not adding any expressive power to the language, is the presence of a language-defined disambiguating mechanism enabling the programmer to give the same name to two routines appearing in the same syntactical context, with the proviso that the argument signature (the number and types of arguments) is different, so that in any call the actual arguments will define which version is intended.

## 3.1 Syntactic overloading: an example

If we have two routines

*open_by_name* (*employee_name*: *STRING*)

*open_by_number* (*employee_id*: *INTEGER*)

(both meant to open the record associated with an employee), a language with overloading allows us to give them a common name such as *open*. The type of the argument determines which of the two versions any call will use: the first version for *open* ("*Jane*") and the second one for *open* (2654).

The Ada language popularized name overloading, which brought a fringe advantage of convenience in a non-OO context (Ada did not introduce OO mechanisms until later versions). Combined with object-oriented programming, syntactic overloading is not only unneeded but harmful, as it breaks the essential simplicity of the OO scheme.

## 3.2 Semantic overloading through polymorphism and dynamic binding

One of the key OO mechanisms is the combination of polymorphism and dynamic binding, which provides a form of *semantic* overloading, far more interesting than the syntactic version. A variable *x* can be polymorphic, meaning that it can at run time denote objects of different types defined by their respective classes (all, however, descendants of a common ancestor in the inheritance hierarchy). Then if some of these classes have different versions of a feature with the same name *f*, dynamic binding determines that the version triggered by any call *x*.*f* (...) is the one corresponding to the type of the object that *x* denotes at that particular time during execution. The standard example, still excellent in spite of having been used so often, is a graphical operation which applies the appropriate version of *display*, different for each kind of figure, where *fig* is polymorphic — dynamically representing a rectangle, or a circle etc., with a specific display algorithm in each case. Polymorphism and dynamic binding are distinct mechanisms, although they go well with each other; many people confuse them. (The functional programming community uses "polymorphism" in a different sense, in the phrase "parametric polymorphism" which denotes a static mechanism, type parameterization, also known as genericity. As already noted, one of the first Eiffel publications [2] discussed its relationship with OO inheritance techniques.)

*fig*.*display*

Unlike syntactic overloading, the combination of polymorphism and dynamic binding truly provides a new level of expressive power, since a run-time mechanism is involved: the selection of the appropriate version of the operation. From a software architecture perspective, it is a natural and necessary extension of information hiding: a client programmer can call an operation such as *display* not only without knowing about its implementation, but also without knowing *which version* of the implementation, among several competing ones, will be triggered in each individual case.

### 3.3 Syntactic overloading is useless and harmful in an OO context

In contrast with the OO combination of polymorphism and dynamic binding, syntactical overloading is a mere notational convenience. The difference between a purely syntactic mechanism and a semantic one (increasing expressive power) is clear if we consider the consequences of *removing* each mechanism:

- If name overloading is not available, the author of a class simply needs to devise a different name — such as the original names above, *open_by_name* and *open_by_number* — for each variant of an operation; then client programmers must use the appropriate name in each call. The burden is light; in fact this solution enforces clarity and helps avoid errors by forcing programmers to specify which version they want. (In the overloaded version, *open* (2654) mistakenly used instead of *open* ("2654") would be accepted but cause wrong behavior, referring to a file identifier rather than a file name.)
- With dynamic binding removed from OO languages, each use of a routine with multiple versions needs to discriminate explicitly between the various cases: if we have a circle, apply the circle algorithm; else, if we have a rectangle, apply the rectangle algorithm; and so on. Every client module using these facilities must know the full list of cases and undergo an update in the case of changes or additions to this list. In contrast, the OO idiom *fig*.*display* with polymorphism and dynamic binding frees clients from having to know which kind of figure *fig* will denote in any particular run-time case, and rely instead on an automatic mechanism to select the right version of the operation in each case.

These observations suggest that syntactic overloading is useless in an OO context. It is, in fact, harmful. In the pursuit of the central issue of programming methodology, maintaining a correspondence between what the program says (the program text) and what the program does (its myriad possible executions), overloading destroys the fundamental simplicity of the notion of class. A class maps a set of operation names to a set of operations. A mathematical model would associate such a mapping with every class:

$$features: Name \nrightarrow\!\!\!\!\rightarrow Feature \qquad \text{-- “}\nrightarrow\!\!\!\!\rightarrow\text{” denotes finite mappings (B.1).}$$

Overloading invalidates this view: *features* is no longer a function (only a relation, see B.1). To obtain the correspondence between names and features, we need an intermediate mapping of programmer-defined names into some internal disambiguated names invisible to programmers. This approach considerably complicates the conceptual framework, with — as seen in section 4 — particularly damaging consequences for the inheritance mechanism.

Even as a notational simplification, overloading often fails in its objective because the disambiguation criterion — the argument signature — is not necessarily the right one, as can be seen in one of the simplest examples that come to mind. To initialize a complex number, 2D-vector or 2D-point, two operations are equally available:

*set_cartesian* (*x*, *y*: *REAL*) -- Initialize by cartesian coordinates. -- /P9/

*set_polar* (*ro*, *theta*: *REAL*) -- Initialize by polar coordinates.

Would it not be nice to use the same name? Unfortunately, the signatures are the same, so — oops — overloading does not work.

Although the need to use different names is, as noted above, at worst a minor nuisance (and at best actually a benefit), overloading causes more than notational trouble because of a crucial language mechanism of C++ and its successors (Java, C#) relies on it.

## 3.4 Overloaded constructors

While it would seem appropriate, when using C++, Java or C#, to stay away from such a fishy mechanism as overloading, these languages do not leave programmers such a choice because they *require* using overloading in the case of "constructors".

The goal of a constructor in a class is to initialize its instances. More precisely (but unknown to most OO programmers, who unless they learn the approach through Eiffel are not taught any formal concepts, let alone DbC) the role of constructors is to make sure that newly created objects satisfy the class invariant. To see how to achieve this goal in a reasonable way, it is useful to look at how Eiffel handles creation:

- A basic creation instruction, for *x* of type *C*, reads **create** *x*. It initializes all fields of the new object to default values (0 for integers and so on as mentioned earlier). This form assumes that *C*'s invariant is trivial since default fields must satisfy it.
- To override the default initialization, the instruction takes the form **create** *x*.*make* (…) where *make* is a *creation procedure* of the class, which performs the needed initializations. "Creation procedure" means the same as "constructor", but the term emphasizes that the operation is a procedure, used here for the particular purpose of initialization.
- For the preceding creation instruction to be valid, *make* must appear in a **create** clause of the class listing its allowed creation procedures. In the point example the clause would list *set_cartesian* and *set_polar*, allowing such creations as **create** *x*.*set_cartesian* (1, 0).
- One of the creation procedures can be *default_create*, a feature of all classes (inherited from the top-level class of the inheritance structure, *ANY*), allowing the simplified form **create** *x*. (A class can redefine *default_create* to override the default initializations.) The absence of a **create** clause is considered a shorthand for a clause of the form **create** *x*.*default_create*.
- A creation procedure is otherwise a procedure like any other; in addition to being used for creation, as in **create** *x*.*set_cartesian* (1, 0), it can also, if it has the appropriate export status, be available for calls, as in *x*.*set_cartesian* (1, 0), which resets an existing object.

- That possibility is available but not always desired. The class author has full flexibility: exporting a procedure such as *set_cartesian* for creation only (it is not available for normal call), for call only, or for both. (Note that the Eiffel export mechanism is fine-grained: you can export features to all classes, or to specific classes and their descendants, or to none at all.)
- The proper way to choose between these variants is the underlying theoretical criterion, unfortunately absent from usual presentations of object-oriented programming: the class invariant. For a procedure (such as *set_cartesian*) to be usable as a creation procedure, its body $b$ must (in a simplified analysis, ignoring arguments) satisfy the Hoare-triple property $\{DEF\}\ b\ \{INV \wedge POST\}$ where $INV$ is the class invariant, $POST$ the procedure's postcondition and $DEF$ the assertion expressing that all fields have their default values. For it to be usable in a normal call, it must satisfy $\{INV \wedge PRE\}\ b\ \{INV \wedge POST\}$ where $PRE$ is the precondition. (The more detailed rule appears in a recent full discussion of the semantics of class invariants [25].) Depending on which of these properties the routine satisfies, it may be usable for creation, for call, or both.

This approach follows directly from the principles of data abstraction and program correctness. In contrast, C++ and its successors have adopted a bizarre set of conventions for object creation. The strangest is the rule that instead of creation procedures, treated as features of the class, creation relies on a set of "constructors" which all have the same name — the name of the class, for example Point. There is no creation instruction, but creation expressions of the form new Point ($arg_1$, $arg_2$, …), which return a newly created object. In the tradition of overloading, the constructors' signatures determine which one is meant in each case.

The absence of a creation instruction forces creation always to repeat the type, even in a typed language environment in which the type of the target (such as the type of *x* in **create** *x* or **create** *x*.*make* (…)) has been declared anyway; it leads to the common Java idiom Point p = new point (…). (Eiffel does have creation expressions too, useful to avoid declaring a variable, for example to create an object and pass it as an argument to a routine in *rout* (***create*** {*C*}.*make* (…)). But to create an object and associate it with a variable there is no need to repeat the variable's declared type.)

The masochistic practice of giving the same name to different constructors makes it impossible to have different constructors of the same structure, as in the frequent Eiffel case of permitting both

```
create point1.make_cartesian (0, 1)
create point2.make_polar (1, Pi / 2)      -- Creating identical points
```

While anyone using 2D-points or complex numbers would want this facility (create a point by giving either cartesian or polar coordinates) as one of the basic mechanisms, it is impossible to obtain in C++, Java or C# since all the constructors would have the same name Point and the same signature (two arguments of float type). The only workaround would be to have a single constructor that handles all cases thanks to a third argument, possibly boolean in this case, indicating which version is desired. Such a contorted solution goes against all principles of good programming.

Programming language sites confirm the impossibility of same-signature constructors, dismissing it in blame-the-victim style. For example, the answer on a Stack Overflow page (bit.ly/3yIdkjY) states that "*there are very few use cases for such an idea even if there was a technically valid way to do it* (*which in Java there isn't*)". A typical cop-out in response to comments about a language's inability to support a programming need: just pronounce that (in Agile development parlance [23]) YAGNI, "You Ain't Gonna Need It"! As the example of cartesian and polar initializations for points illustrates, the need is normal and reasonable. What is unreasonable and unjustifiable is the C++/Java reliance on a single name to cause ambiguity, and an artificial criterion, signatures, to address the resulting self-imposed predicament (when at all possible).

Just as unreasonable is the resulting damage to program clarity: to understand what a particular creation expression new C (x, y, z) means, the client programmer has to go to the text of the class and peruse the list of constructors to see which one has a signature matching the types of x, y, z. Also note that inheritance-based subtyping can make the disambiguation rules complex.

What is most puzzling here is the absence of any need, or justification, for using overloading. Even the most unexperienced parents know to give their children different names. (While many families have a daughter called Jill, it is not wise to give that name to two of your children.) In programming too, different operations appearing in the same class should have different names, so that calls (such as those in creations using *make_cartesian* and *make_polar* above) are clearly differentiated. (Of course, different classes — like different families each calling *one* daughter Jill — can use the same feature names, opening the way to the truly useful mechanism of dynamic binding.)

By breaking the fundamental simplicity of a class as defining an injective mapping in $Name \nrightarrow\!\!\!\!\rightarrow Feature$, overloading also jeopardizes, as we will now see, the fundamental OO mechanism of inheritance.

# 4 INHERITANCE AND INTERFACES

Inheritance is another area where a simple approach, directly following data abstraction principles and seeking simplicity (including simplicity of the mathematical model) removes accumulated sediments of language obfuscation.

Misconceptions often arise because of an unduly restrictive view of inheritance. An earlier article [10] explored the wealth of applications of this technique.

## 4.1 Redefinition

In the OO mechanism of inheritance, a class *D* may be defined as inheriting from a class *C*, to express that *D* has all the features and invariant clauses of *C*, in addition to any that it may define for itself.

Fundamentally for the expressive power of OO design and its approach to reusability, it is also possible for *D* to redefine, or "override", a feature inherited from *C*, by giving it a new implementation. For example a class *RECTANGLE* retains the function *perimeter* from its parent *POLYGON* but can provide a better implementation, specific to polygons that are rectangles.

If the single-mapping property holds (and hence no overloading is permitted), the situation is simple and the mechanism easy to explain: the presence of a new feature with the same name as an inherited one (and a conforming signature) means that it overrides it, in the sense of the overriding union operator $\Cup$ (B.2). To avoid any uncertainty or confusion, Eiffel requires the inheritance clause (the place where *D* states it inherits from *C*) to specify **redefine** *f*, *g*, … for redefined features. Unless the list includes all redefined features, the class is invalid since it includes different features with the same name.

## 4.2 Renaming

Another useful facility is renaming, which enables a descendant to inherit a feature under a different name.

The mathematical meaning is also simple: if *m*, an element of $Name \twoheadrightarrow Feature$, is the names-to-features mapping defined by *C* (3.3), and *D* performs a set of renamings, mathematically a finite permutation $\sigma$ (defined in B.3), then the new name-feature mapping associated with *D* is $\sigma^{-1}\ ;\ m$. As $\sigma^{-1}$ is injective per the definition, no overloading will result.

Comparing renaming and redefinition is illuminating, and combining them useful. Renaming changes the name, without changing the feature. Redefining changes the feature, without changing its name. You may need one of these operation, or the other, or both ($(\sigma^{-1}\ ;\ m) \Cup m'$).

It is a source of constant wonder that programmers and students do not understand the fundamental simplicity of these ideas, and imagine all kinds of complicated scenarios, failing to understand the basic distinction between a feature and its name. They are of course not to blame, but have fallen victim to approaches that turned simple matters into complicated ones.

## 4.3 Multiple inheritance

Inheritance supports a refinement process whereby we build new abstractions on the basis of existing ones. It is an essential mechanism of object-oriented programming, applying to program construction the basic scientific concept of taxonomy.

Often, we need to extend or specialize not only one existing abstraction but several. In a hierarchical windowing system, a window is both a graphical object and a node in a tree. It has all the prerogatives (in the sense of features and semantic properties including invariants) of both of these notions. In such examples, which constantly arise in the practice of object-oriented programming, multiple inheritance is the clear solution, offering an advanced form of reuse.

Unfortunately, a botched design for multiple inheritance in C++ gave this technique a bad name. The first culprit is, once again, overloading. In the simple overloading-free world of OO programming, two classes can only be combined through multiple inheritance if their feature names do not clash. If a name clash does initially arise, it is easy to resolve: just use renaming. So if both *A* and *B* have a feature called *f*, we may define, in Eiffel

```
class D inherit
    A rename f as A_f end          -- Obviously it would be enough to
    B rename f as B_f end          -- rename one of the two.
feature
    ... New features of D ...
end
```

With overloading, both inherited versions can coexist if their signatures are different. No wonder that with such complicated semantics multiple inheritance gets a bad rap.

This area is one where wrong preconceptions are so entrenched that discussions become impossible, so convinced are programmers (and even beginning students) that this simple matter has to be utterly complex. Those trained in C++ immediately go into discussions of data structures — "Vtables" and other tables of pointers, techniques that were in fact pioneered by Eiffel implementations but should be the concern of the small community of compiler writers, not general programmers (who in Eiffel are thankfully spared from having to know them). A little as if we had to explain conditional instructions by talking about Intel's *jne* and *jnz* CPU instructions. All that the Eiffel programmer needs to know is that multiple inheritance causes no loss of run-time performance (finding the right version of f in a call *x*.*f* (...) always takes constant time) thanks to techniques invented at the time of the very first Eiffel compilers.

As to Java/C# programmers, they have already been instructed that multiple inheritance is bad (because C++ could not get it right) and "interfaces" solve the problem. They do not, by a long shot.

## 4.4 Interfaces

Early attempts to introduce principles of data abstraction and information hiding into programming languages led (even though they were posterior to the invention of OO programming in Simula 67) to modular mechanisms that required a module to be defined in two separate parts: the interface (or "specification") part and the implementation (or "body") part. The interface part only includes the list of features (taken here in the general, non-necessarily-OO sense of functionality elements) with their signatures; the implementation contains the actual algorithms and data representation choices. The typ-ical "*modular languages*" applying this idea are Modula-2 and Ada. (It may be noted that the C language, while not providing an actual module construct beyond the concept of function, already promoted a similar discipline by directing programmers to define every program component *prog* through two files, *prog*.*h* for its functions' interfaces and *prog*.*c* for their implementation.)

## 4.5 Limitations of interfaces in pre-OO modular languages

While historically a healthy step towards proper program structuring, the interface/implementation separation of modular languages suffers from important limitations and defeats modern design methodology.

The most obvious deficiency is that these interfaces are not real specifications since they are purely structural, specifying only the names and signatures of the features and saying nothing of their semantics. A more useful notion of interface would have to include Design by Contract mechanisms, specifying semantic properties — preconditions and postconditions of features, type invariants — that would be binding on the implementations. Without these properties, the specification given by an interface in C, Ada, Modula-2 (as well as Java and C#) does not provide client programmers with what they need to use the module in accord with the principle information hiding: enough documentation to use the functionality without knowing its implementation.

There is an even worse problem with the interface-implementation duopoly. A dogmatic split into just two categories misses the need for proper models to include an extended chain from the most abstract descriptions to fully worked-out implementations. Here inheritance — aided by the mechanisms for adapting contracts, making them more general or more specific as explained in 5.4 below — leaves the interface/implementation concept in the dust.

## 4.6 Programs with holes

A brilliant invention of OO programming is the use of inheritance to provide a full spectrum of refinements, through the technique called "programs with holes" in [4] and [11]. It reflects a basic difference between definitions in mathematics and programming.

Consider the notion of a total order relation:

- In mathematics, we can define it by specifying that a certain relation, "$<$", satisfies certain properties (transitive, irreflexive, total). We can then deduce other properties, for example that it is also asymmetric, and that it has variants "$\leq$", "$>$", "$\geq$" with their own properties (for example "$\leq$" is transitive, symmetric, reflexive and total). We could in fact define the total order concept from any of them, but for the definition we only use properties that are strictly necessary (those of "$<$"); all properties of the other relations are theorems, not part of the definition but consequences of it.

- To define the corresponding structure in programming, the perspective is different. The four operations are equally important as part of the interface; there is no reason to divide them into definitions and consequences.

In a corresponding class *COMPARABLE*, we may still choose one as unimplemented, or "deferred", and the others "effective", that is to say implemented. For example:

```
lesser alias "<" (other: COMPARABLE): BOOLEAN
        -- Is current object strictly less than other?
    deferred end

greater alias ">" (other: COMPARABLE): BOOLEAN
        -- Is current object strictly greater than other?
    do
        Result := other < Current
    ensure
        Result = (other < Current)
    end
   … Similarly for "≤" and "≥" …
```

(The **alias** clauses make it possible to call the features in infix form, for example $a < b$ as a synonym for *a*.*lesser* (*b*).) Feature *lesser* is deferred, but all the others are effective, defined in terms of it ("**do**" followed by an implementation, instead of "**deferred**").

Class *COMPARABLE* is deferred, as is the case with any class that has at least one deferred feature. But it is not *fully* deferred: some of its features, such as *greater* and others cited above, are effective, defined in terms of it. *COMPARABLE* is a "**class with holes**": the implementation of *lesser* is left open, a "hole" for descendants to fill, but the others are already filled. A descendant class will effect (implement) *lesser*, as in

```
class STRING inherit COMPARABLE feature
    lesser alias "<" (other: COMPARABLE): BOOLEAN
            -- Is current string strictly less than other?
        do … Implementation of lexicographic order on strings … end
    …
end
```

Key here is the property that such classes do not need to, and should not, do anything about the other ordering features, "≤", ">" and "≥", which were defined in an effective form in the general class, *COMPARABLE*, and remain unchanged, although of course their implementation depended on "<" and follows its own implementation in *STRING*.

## 4.7 Combining abstractions

Such a *COMPARABLE* class exists in the Eiffel Kernel Library, which in a similar way includes a class *NUMERIC* describing number-like objects; more formally, members of a commutative ring, with operations "+", "–", "∗" and "/" and elements zero and one. Like *COMPARABLE*, *NUMERIC* may define effective features based on the deferred ones, for example *squared* in terms of "∗".

To define classes such as *INTEGER* and *REAL*, it suffices to use multiple inheritance from *COMPARABLE* and *NUMERIC*. As in other cases of inheriting from these classes, the effective classes provide implementations of the features inherited in deferred form, such as "<" and "+" — but those only, since the effective ones automatically follow.

In Java and C#, classes are limited to single inheritance. It is a widely held *belief* (the word seems the most appropriate) that this limitation does not matter because a class can inherit from multiple interfaces. But as examples such as the above show, this possibility is a lame imitation of OO techniques. If a class conceptually needs to inherit from two others, it will have to use a kludge replacing one of them by an interface. If *COMPARABLE*, for example, becomes an interface:

- Features that should normally be declared as effective ("≤", ">" and "≥") must be declared with their signature only, as if they were deferred.
- They will have to be implemented in every descendant class, like "<". But unlike with "<" those implementations are always the same (as given for ">" above).
- Somehow the interface must then come with a kind of user's manual: dear author of a descendant, if you implement the interface, please, please, always implement these features in a specified way.
- There is no guarantee that programmers of descendant classes will follow that injunction. Errors will creep in if authors of descendant classes miss it and introduce implementations that are incompatible with "≤".
- Contracts could alleviate the problem: in Eiffel, you can give ">" (for example) a postcondition stating **Result** = *other* < *Current* (I am greater than you only if you are lesser than me). The rules on contract inheritance make such constraints binding on descendants. (In Eiffel they open up the possibility of actually redefining the implementation for a routine like ">" from "<" in *COMPARABLE*, if desirable for performance or any other reasons, while preserving the correct semantics.) But most languages do not have contract mechanisms.

## 4.8 Interfaces as automatically derived artifacts

While interfaces in the Ada or Java sense are not a useful *software construction* tool, the need remains for a way of *documenting* existing classes (deferred or effective) abstractly, without implementation details. Interfaces cause some code duplication: the abstract information — names of features, their type signatures, header comments, contracts if supported by the language — is repeated in the concrete module as well. A better approach is to write all elements (abstract and concrete ) only once, as repetition in software is always bad, and to *extract* the abstract elements from the text, for purposes of documentation.

The EiffelStudio environment provides that functionality through the notion of *views*, in particular the **contract view**, which extracts the elements just cited (name, signature etc.) from a class text, leaving out all routine bodies and all non-exported (private) features. In accordance with the Uniform Access principle (section 1, particularly 1.3), functions and attributes appear in the same form in that view.

The result is an effective form of documentation, abstract but guaranteed to be in sync with the actual code since it is extracted from it. (One of the worst problems with documentation arises when the software evolves and its documentation is no longer accurate.) It can be produced in various formats, including HTML and PDF.

The contract view is one of several produced by the environment. The **flat view** is a reconstructed form of the class text including all inherited features, and applying all renaming and redefinition clauses. The **interface view** is the contract view of the flat view; as the name suggests, it provides the interface documentation of the class, as useful to client programmers (who do not need to distinguish between features introduced in the class itself and those inherited from ancestors). Like a Java interface — but produced by tools, not programmers. As an example, here is the extract of the interface view for the library class *LINKED_LIST*, providing the specification of a specific routine:

```
copy (other: like Current)
        -- Update current object using fields of object attached
        -- to `other`, so as to yield equal objects.
    require -- from ANY
        other_not_void: other /= Void
        type_identity: same_type (other)
    ensure -- from ANY
        is_equal: Current ~ other
```

Note the inherited assertions, with a comment (added by the tool) indicating the class where they originated, per the rules of assertion inheritance covered in 5.4 below.

# 5 CONTRACT MECHANISMS

Design by Contract is the language trait that most people who have heard of Eiffel can cite. It is also among the most misunderstood, for example by being assimilated to defensive programming, its exact opposite.

Under the assumption that the reader has heard of preconditions, postconditions, class invariants, loop invariants and loop variants, this discussion will not explain the basic concepts. It limits itself to dispelling a few common misconceptions.

## 5.1 A close integration with the language

The first mistake is the often heard comment that "*one can use Design by Contract in any language*" (often intended, in the author's personal experience of many years, as a compliment, as in "*I like your ideas, but I apply them in …*", to which the only reasonable reaction is to smile politely without mentioning the sheer absurdity of the idea). The evidence massively disproves this view: while dozens — probably hundreds — of DbC extensions have been proposed for just about every significant programming language, almost none has gained wide acceptance. The only exceptions are Ada Spark and JML, which have achieved some usage in the important but narrow field of formally verified mission-critical applications (and largely, in the second case, for research).

Why? We may immediately dismiss a folk explanation: that the concepts are too hard for most programmers. This old canard has not basis in fact. Contracts are boolean conditions. Anyone who can write an **if** … **then** … **else** … conditional — that is to say, any programmer — understands this notion and can use it for specification just as well as for implementation. The long experience of many educational institutions in teaching the concepts, including fourteen consecutive years of the introductory programming course at ETH Zurich by the author and the resulting textbook [17], are evidence enough.

The true reason is simple: Design by Contract can only work if tightly integrated with the fabric of the language and the development environment (IDE). For example, the rules about inheritance specify that the precondition of a redeclared routine must be kept or weakened, and its postcondition kept or strengthened. (They are commonly attributed to the "Liskov Substitution Principle" although they come from much earlier Eiffel publications from 1986-1988 [1][3][4].) The Eiffel mechanism enforces this methodological rule by prohibiting a redeclared version of a routine from using the normal **require** or **ensure**, permitting only no clause — in which case the routine retains the original pre- or postcondition — or clauses of the forms **require else** *new* and **ensure then** *new*, which respectively "or" and "and" the *new* part with the original, automatically yielding a weaker resp. stronger assertion. Without this kind of built-in language rule, the use of pre- and postconditions is largely pointless since you have to duplicate the original assertion manually — a chore that no programmer will tolerate for long — or risk inconsistencies.

## 5.2 Class invariants

Contracts without built-in language support become even more unrealistic in the case of class invariants. Dating back to an article by Hoare in 1972, class invariants (see [25] for a recent review of the concept) can and should play a major role in software design by expressing the consistency conditions that underlie any significant class. But a class invariant can only be expressed at the level of a class.

Without a dedicated language construct, the invariant has to be repeated (causing massive code duplication) as an addition to both the precondition and the postcondition of every exported routine, and its every redeclaration in a descendant! We are bordering on madness here, and of course no one, including people who claim to "*apply the Design by Contract methodology but in an ordinary language*", uses these facilities, crucial to obtaining any benefit from DbC.

## 5.3 A plethora of run-time checks for free

Here is a concrete example of these advantages. In section 7.3 below we will see the full version /P19/ of a class invariant from an actual program by the author, of the form

$$\forall\, x: A \mid \forall\, y: H(x) \mid \forall\, z: L(y) \mid some_property(z)$$

where *A* is an array of hash tables *H* (*x*) (for every element *x* of the array) of lists *L* (*y*) (for every element *y* of such a table). One of the applications of contracts is the ability, provided by the compiler and IDE, to evaluate assertions as part of the testing and debugging process. The invariant clause expresses an elaborate consistency condition affecting a data structure which in practical application may total tens of thousands of elements or more. Such contract elements are in fact easy for the programmer to write, since they just express the assumptions that preside over the writing of the code. In normal development — even without the prospect of program proofs — every execution of an exported routine of the class will cause the conditions to be evaluated, resulting in an enormous amount of automatic consistency checks. Once runs of the code proceed without violating any of these conditions, the programmer's confidence in the code, while not absolute (see the discussion of proofs below), is higher than any traditional approach to testing can provide.

## 5.4 At the very core of inheritance concepts

Design by Contract specifies that the invariant of any class includes, in the sense of logical "and", the invariants of its parents under inheritance. Like the other rules governing inheritance (weakening/strengthening of pre/post-conditions noted in 5.1 above), applying this principle requires language support; no manual or tool-based addition to a non-contract-equipped language can provide a usable substitute.

These inheritance rules of DbC are not just interesting theory but part of the very definition of inheritance:

- Without the "and"-ing of invariants, inheritance is just a way to reuse some facilities, little more than listing an "include file" in C. The invariant rule shows what inheritance truly is: the "is-a" (subtyping) relation, not just providing the descendants with the ancestor's features but also constraining them by the same semantic rules as these ancestors.
- The weakening/strengthening rules are similarly critical to understanding the fundamental OO tools of polymorphism and dynamic binding. If you redefine *r* in a descendant class, meaning that *x*.*r* (...) will trigger a different operation depending on the type of *x* (ancestor or descendant) in any particular execution, you have a very powerful mechanism, but also a dangerous one: what prevents a descendant class from redefining *r* in a way that violates the original intent? The only way to answer that question is to define that original intent and enforce it in descendants; in other words, use a precondition and a postcondition, and rely on the weakening/strengthening rules, which mean that the new version can do better — be more tolerant of the input, or produce better output — but can never do worse.

This approach makes it possible to teach inheritance properly. One can only regret that the vast majority of OO programmers have never heard of these concepts and hence lack methodological guidance to use inheritance, a powerful but risky tool of object technology.

## 5.5 Understanding the Design by Contract methodology

In Eiffel, the use of contracts is neither abstract advice nor a burden, but a tool for building software that will be correct by construction.

People who heard about the ideas but did not study them often confuse them with "defensive programming", their opposite. Defensive programming is the practice of protecting operations, through and through, with checks of their applicability. The more checks, the safer the software (supposedly).

Contracts are entirely different. The idea is, as with a human contract, to *distribute* the responsibility for consistency conditions between the various parties in the myriad interactions that take place between software elements — clients and suppliers in a system. Distribute, but never replicate. If negative and non-negative values are to be handled differently for an operation, then either:

- The operation (the supplier) is responsible: it processes all values, producing different results in the two cases. The postcondition specifies the results in these various cases.
- The clients are responsible: for example the routine accepts non-negative values only, and puts that requirement in its precondition.

Defensive programming would encourage checking on both sides. This approach is haphazard and downright dangerous. If everyone is responsible, no one is responsible; errors will creep in anyway. There is no specification (in the form of a contract), so it is easy for each of the parties to believe that the other is in charge of a particular condition.

(On this particular topic, ample empirical evidence exists. For many years we conducted the Distributed Software Laboratory course, also known as DOSE, involving a cooperative software development project conducted by student teams from universities in many countries around the world. We closely monitored student activity and systematically collected project information. In one of the first iterations we "encouraged" students to use contracts. Most — being new to Eiffel — did not, and the result was buggy software, suffering from a plethora of conditions left unchecked out of the assumption that the other team was responsible. From the second year on we *mandated* the systematic use of Eiffel contracts, actively enforcing the mandate when reviewing code commits, and these problems disappeared entirely. Several publications present the empirical results; see [20] and the site URL given in that reference.)

Another problem with defensive programming is that it adds code, often redundant. Code bloat is not the solution, but only compounds the problem. It may introduce further bugs. In fact, redundant code is more likely to contain bugs since it typically does not get exercised during testing. Duplication bugs can be particularly subtle, as the conditions checked redundantly may be similar but not be identical, particularly after the software undergoes change. Contracts involve no duplication or redundancy: they *complement* the code, which expresses the "how", by stating the intent, the "what" (sometimes even the "why"). That is the reason why run-time contract monitoring (5.3) is effective as a testing technique: it compares reality to intent, expressed at a different level of abstraction.

A good criterion for finding out if a team is pretending to use Design by Contract but in reality just adding layers of (potentially noxious) defensive code is to ask if routines check their preconditions (as in a square root routine requiring a non-negative argument $x$ in its precondition, and *also* including **if** $x < 0$ **then** … in its code). With Design by Contract this sloppy defensive practice is never acceptable. (As the author was once visiting Redmond, a Microsoft Research team working on the "Code Contracts" library proudly asked to meet him to present its work. A quick examination showed that the body of routines with preconditions — for example, that a queue is not empty when you want to get its first element — *tested* for these conditions, raising exceptions when they did not hold. Time to put a courteous end to a pleasant encounter.)

Any redundancy useful for testing and debugging will be provided not by duplicating code but through the automatic tools of run-time contract monitoring.

The underlying view of programming methodology is that professional, correctness-by-construction software development does not consist of blindly throwing extra "just in case" consistency checks all over the code (which one can compare to adding props all over a structurally deficient building), but of identifying the consistency conditions that the computation requires, assigning the responsibility for each of them (on the basis of architectural considerations) to *one* of the parties, and sticking to the decision.

## 5.6 An integral part of the approach

What makes Design by Contract practical is its applicability at many levels, supported by all elements of the Eiffel methodology, language and IDE (EiffelStudio):

- In the methodology, contracts provide a strong guidance to the programmer, a form of correctness by construction.
- The language closely integrates the concept, woven into the rest of its mechanisms, particularly inheritance as just seen.
- Contracts provide the basic documentation tool, in the form of "contract views" and their variants (4.8), produced by the environment.
- The IDE also supports run-time assertion monitoring, a key tool for testing and debugging as discussed above. Modern testing frameworks of the "xUnit" style also rely on assertions as test oracles, but the assertions are written only for testing. Eiffel's assertions are an integral part of the program, written with it, not as an afterthought. They express essential semantic properties of the code (preconditions, postconditions, invariants), not just testing oracles, and serve software quality goals. Note that an extensive effort around the AutoTest framework [18] has taken advantage of built-in contracts to extend xUnit ideas and, in particular, generate tests automatically.

For verification, run-time contract monitoring is a powerful technique, but one would ideally expect *static* proofs of correctness (defined as the conformance of the implementation to the contracts). The AutoProof program proving environment for Eiffel [22] pursues this goal. While still a research tool, it has reached significant capabilities — which anyone can try out in a browser at autoproof.sit.org — and has been used (in Nadia Polikarpova's ETH PhD thesis, se.ethz.ch/people/polikarpova/thesis.pdf) to prove formally and mechanically the correctness of a full library of fundamental data structures and algorithms, EiffelBase 2.

# 6 STATIC CLASSES AND SHARED INFORMATION

Object-oriented programming is the quintessentially modular method: it organizes systems into moderately sized units (classes), each built around a well-defined abstraction. The decentralization of the resulting software architecture is key to obtaining the method's benefits of extendibility (ease of adapting software to changing needs or circumstances) and reusability (avoidance of repetitive or duplicated programming work).

Not everything can be decentralized: even the most modular architecture retains the need for some shared information. The most elementary example is console output: all parts of a program may need to write to a common medium.

## 6.1 Class methods and static classes

Smalltalk addressed this need with "class methods", a technique that assumes that classes can at run time be treated as objects. This approach leads to interesting conceptual developments ("Metaobject protocols") but seems like overkill for solving a simple need. C++ and its successors introduced a concept specially designed for that need: static classes, which cannot be instantiated.

## 6.2 Once routines

The principal issue with shared information is its initialization. A class — a normal one, not "static" — describes a set of potential run-time objects, and specifies mechanisms to initialize them: creation procedures or constructors (discussed earlier in section 3). But with a static class there are no associated objects.

Eiffel has an original solution: once routines. If the body of a routine starts with the keyword "**once**" instead of "**do**", it will be executed not for every call to the routine, but only for the first one. In the case of a procedure, we can use this mechanism ito initialize a shared resource on-demand, without having to know in advance which caller — if any

— will first need it. For example various routines of a graphical library may require that the display has been set up. The library can include a procedure

```
setup_display                                        -- /P10/
    once
        … Operations to set up the display …
    ensure
        is_initialized
    end
```

Then the routines that need a set-up display start by calling *setup_display*. Only the first to do so will actually cause the setup operation; subsequent calls do nothing.

In the case of a function, the result will be computed the first time around, and returned again by every subsequent call. This facility works well for shared objects, as in

```
Console: IO_MEDIUM                                   -- /P11/
    once
        … Preparatory operations …
            create Result.make (…)
    end
```

Note the use of a creation instruction of target **Result**, typical of this design pattern. The first call creates the console; all subsequent ones use the same object. It does not matter who issues that first call. If no element of the software calls *Console*, the object will not be created; this property is important as a system may potentially require many such shared objects, but it would be a waste of resources to create all of them in all cases. In addition, since they are created individually and only on-demand when first needed, they avoid delaying system startup through the creation and initialization of a whole set of objects.

## 6.3 Instance-free features

In an object-oriented architecture where every element of information appears in a class, where should shared information, such as once functions, appear?

One solution is simply to use inheritance. A class gathering shared information can serve as parent for any other class that needs to use the corresponding features. It is common to frown on this use of inheritance, but it is legitimate [10] since such classes embody valid abstractions; the real reason for criticism is the absence of usable multiple inheritance in most OO languages (section 4). An alternative to inheritance in this case is, however, available: qualify the feature with the class name, as in

```
{IO}.Console.write ("message")                       -- /P12/
```

where *IO* is the class containing the above declaration of console. *{IO}.Console* is known as a "non-object call" since it does not involve an object, only a class.

It is valid to use a feature (here *Console*) in a non-object call only if it is *instance-free*, meaning that its implementation only depends on the properties of the class as a whole and not of any specific instance. A class feature may not refer to any attributes, except for constant attributes (declared in the form *Pi*: *REAL* = 3.1415926535).

Since these properties characterize the implementation, they do not figure in the interface view (4.8). To be sure that a non-object call such as /P12/ will be valid, however, programmers must know that the feature, here *Console* from class *IO*, is instance-free. The rule is that it must have a postcondition (part of its **ensure** clause) of the form

*instance_free*: *class*

which does show up in the interface view. That clause itself is valid if and only if the feature satisfies the instance-free requirement.

### 6.4 Varieties of once routines

The meaning of “once” is subject to parameterization using a “once key”. By default **once** is a synonym for **once** ("*PROCESS*"), but in a multithreaded environment you can use **once** ("*THREAD*").

Another available choice is **once** ("*OBJECT*"). A routine thus specified will be executed on its first call, if any, on any given instance of the class. As an example, an object representing a currency may have has an associated history — the list of its exchange rates over past years — in the database. The corresponding *history* query should not be an attribute: all histories of all currencies would be loaded in memory at all times. But it should also not be a normal **do** function, since it would be executed (loading all relevant history values) each time it is needed, even though the result is always the same. The proper implementation is as a once-per-object function, which achieves the proper tradeoff between attributes and functions (reinforcing once again the concept of query and the Uniform Access principle). If a particular object never needs its history, the history will not be loaded. If it requires it once or many times, it will be automatically loaded the first time around, then kept in the object structure for the rest of the execution.

## 7 OLD SYNTAX FOR NEW SEMANTICS

The example of assigner commands and assigner calls (1.5) illustrates a guiding principle for the syntax of Eiffel: maintain the integrity of the methodology, including design principles (semantics), while accommodating well-established notations (syntax).

This approach stands in contrast to the solutions taken by other OO languages, which sometimes depart from OO principles to retain compatibility with older approaches. In C++, for example, basic types such as integers are not part of the OO type system but remain special types with special properties. Java offers both OO versions (with lesser performance) and traditional versions, with a tricky process of going back-and-forth between them. Eiffel uses instead a consistent OO-based type system. Every type, basic or not, is defined from a class. Classes for basic types are part of the basic library and can be brought up in the EiffelStudio environment, where their display starts like this:

```
frozen expanded class INTEGER_16 inherit

    INTEGER_16_REF
        redefine
            is_less,
            plus,
            minus,
            product,
            quotient,
            power,
            integer_quotient,
            integer_remainder,
            opposite,
            identity,
            as_natural_8,
            as_natural_16,
            as_natural_32,
            as_natural_64,
            as_integer_8,
            as_integer_16,
            as_integer_32,
            as_integer_64,
```

(different sizes are provided but you can use just *INTEGER*), with features such as

```
plus alias "+" (other: INTEGER_16): INTEGER_16
        -- Sum with `other'
    external
        "built_in"
    end

minus alias "-" alias "−" (other: INTEGER_16): INTEGER_16
        -- Result of subtracting `other'
    external
        "built_in"
    end

product alias "*" alias "×" (other: INTEGER_16): INTEGER_16
        -- Product by `other'
    external
        "built_in"
    end
```

and class invariants expressing basic mathematical properties. These classes are only "special" in that you cannot modify them (see the "built-in" mentions) and compilers handle them specially, allowing the same run-time efficiency as in a language like C where they directly correspond to machine operations. (Automatic conversions, for example from integers to reals, also rely on a general mechanism, not special rules; see 7.4 below.)

All operations conform to the OO model of computation: an addition is, conceptually, an OO function call of the form *a*.*plus* (*b*), and can be written that way. Such syntax is not, of course, what most people (at least most people who learned arithmetic before object-oriented programming) want to write; but then the **alias** "+" specification in the declaration of *plus* makes it possible, as explained earlier, to write *a* + *b* as a synonym. Not a step out of OO programming but a different syntax for the semantics of calling a function on an object (which the compiler can translate into plain arithmetic-addition code).

Many Eiffel mechanisms follow this spirit of providing convenient syntax, often reflecting traditional modes of expression, for mechanisms that fully respect the object-oriented paradigm. A few more examples follow.

## 7.1 Aliases of various kinds

The alias mechanism makes it possible to define prefix or infix equivalents for functions with (respectively) 0 or 1 argument. There may be more than one alias, in particular to accommodate Unicode operators, as in

```
quotient alias "/" alias "÷" (other: INTEGER_16): REAL_64
        -- Division by `other'
    external
        "built_in"
    end
```

(To insert Unicode characters such as "÷" with an ordinary keyboard it suffices to use a pull-down menu in the IDE.) Specific aliases include the *bracket* alias, enabling standard array notation. For example the basic element-access feature for arrays is declared as

*item alias* "[]" (*i*: *INTEGER*): *G* **assign** *put* -- /P13/

making it possible for an array *a* and an integer *i* to use *a* [*i*] as a synonym for *a*.*item* (*i*), which yields the element of index *i* in *a*. Thanks to the **assign** part, the notation *a* [*i*] := *x* is permitted, as a synonym for *a*.*put* (*x*, *i*), which updates the entry. The instruction

*a* [*i*] := *a* [*i*] + 1

is easier to read than its default-OO-style equivalent *a*.*put* (*a*.*item* (*i*) + 1, *i*) (let alone the version using *plus* instead of "+").

The use of *a* [*i*] on the left side relies on the assigner command mechanism of 1.5 (here we specify *put* as an assigner command for *item* in /P13/).

In these and other facilities a general principle of the language's design is that whenever a simplification mechanism is available to language designers it is also offered (if potentially useful) to all programmers — class designers — as well.

Allowing *a* [*i*] as a shortcut for *a*.*item* (*i*) is not a special facility for placating programmers used to traditional array syntax but a general mechanism that any class can use. For example a hash table class (such as *HASH_TABLE* in the EiffelBase library) can define a bracket alias for the operation that accesses an element through its key, and associate it with the appropriate assigner command, allowing the writing of an insertion as

*age* ["*Jill Smith*"] := 29

## 7.2 Once classes

Another example of making traditional concepts fit in the OO methodology is the concept of once classes. It covers a generalized form of the "enumeration type" idea. For a long time Eiffel only had a basic enumeration type mechanism (a specific use of integers), based on the methodological observation that choice between specified variants is best handled through inheritance, polymorphism and dynamic binding — a YAGNI-style attitude (3.4). The need for a specific mechanism did impose itself in the end. (Since this article often criticizes Java, it is fair to note that this specific mechanism was influenced in part by one present in that language.)

A class declared **once class** has all the possibilities of a normal class, but its creation procedures must all be once routines. They yield specific instances, each with specific values of the attributes. (Think of a class *US_STATE* with 50 such creation procedures.) It is then possible to discriminate between instances, using an **inspect** multi-branch, or iterate over some or all of them using an **across** loop as will now be described.

## 7.3 Iterators

Many languages have special constructs for iterating over specific data structures. Eiffel conceptually has a single kind of loop:

**from** *Init* **invariant** *Inv* **until** *Cond* **loop** *Body* **variant** *var* **end** -- /P14/

(where the **invariant** and **variant** clause, and they only, are optional). It can serve to iterate over a data structure *struct* to which a cursor *curs* has been associated:

**from** *curs*.*start* **until** *curs*.*after* **loop** -- /P15/

*some_operation* (*curs*.*item*) ; *curs*.*forth*

**end**

with operations on cursors as illustrated below. (In early Eiffel-related literature [4] [8][11] these were operations on *struct* itself, using its own built-in internal cursor, but it is better, in particular in a concurrent programming context, to use an external cursor, so that various clients using a data structure can each maintain their own cursors, keeping different views of that same structure.)

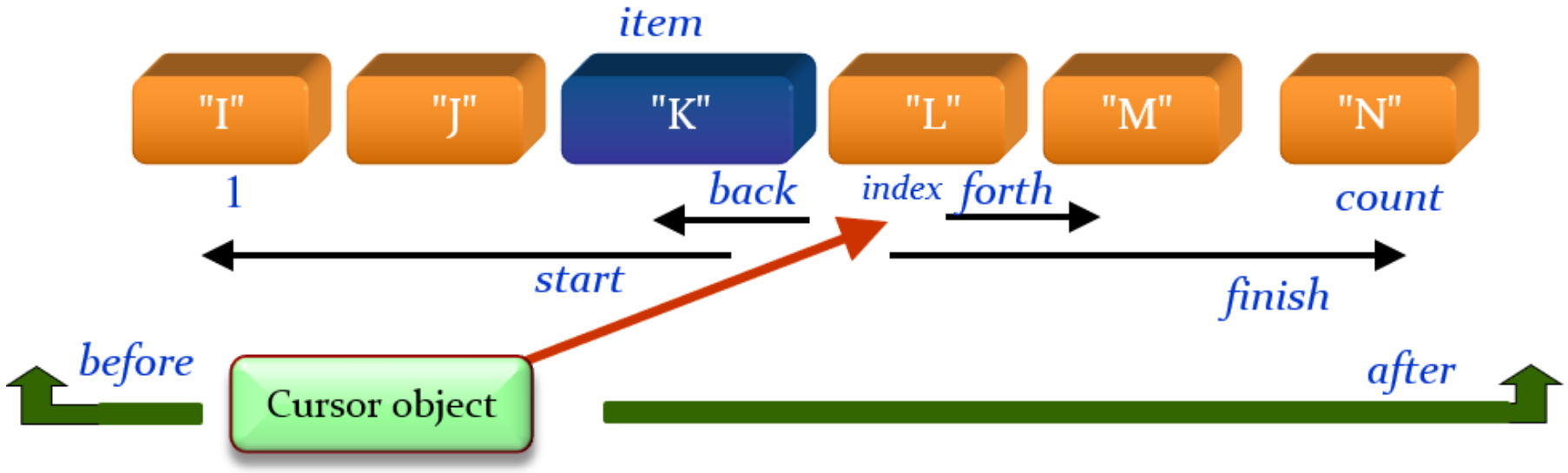

/P15/ assumes that the class for the structure under iteration inherits from the library class *LINEAR*, with an associated class for the cursor. Many usual classes such as *ARRAY* and list classes are descendants of *LINEAR*. Any programmer-defined class can also inherit from it, defining its own implementations of features *start*, *before*, *item* etc. /P15/ is common enough to justify an abbreviation hiding the cursor manipulation details:

**across** *struct* **as** *x* **loop** *some_operation* (*x*) **end** -- /P16/

The name *x* serves to denote the current element, like the bound variable *i* in an expression of the form $\sum_{i:\,1..n} p\,(i)$ in ordinary mathematical notation, or (in both math and Eiffel)

$\forall\ i: E \mid property\ (i)$ -- /P17/

Quantified expressions such as /P17/ are commonly needed, particularly in assertions (contracts). /P17/ can be written exactly as given, as well as its counterpart using ∃. In line with previous ideas, /P17/ is an abbreviation for the keyword-based form

**across** *struct* **as** *x* **all** *property* (*x*) **end** -- /P18/

It is interesting to note the power of expression that these mechanisms provide. As previewed in 5.3, a class invariant clause in a program recently written by the author, handling a complex data structure — an array of hash tables, with integer keys, of linked lists of triples of integers — reads (here with the original identifiers and comments):

```
                                                                  -- /P19/
∀ httpl: triple_successors ¦          -- httpl.item: hash table of list of triples
        ∀ tpl: httpl ¦                -- tpl: list of triples (on the other hand,
                                      -- @tpl.key is a key (a tag) for the hash table)
            ∀ tp: tpl ¦               -- tp: triple
               tp.tag = @tpl.key ∧ tp.source = @httpl.cursor_index
```

("∧" is defined as an alias for **and** in the library class *BOOLEAN*. For an iteration variable *c* denoting a cursor, *@c* denotes its index in the iteration: 1 when the cursor is on the first element and so on.) The benefit of such powerful assertions for extensive testing was mentioned earlier (5.3). What is not possible in ordinary programming languages is the ability to use first-order logic in classical notation.

## 7.4 Conversions

Another application of the principle of providing general mechanisms rather than ad hoc solutions is the conversion facility. In typed languages, conversions sit somewhat outside of the type system mentioned in section A.5. Most languages allow the assignment *r* := *i* for a real target *r* and an integer source *i*, causing a conversion; such cases, however, do not fit in the general type system, being instead treated as special.

Conversions can be useful for many types, not just predefined ones such as *REAL* and *INTEGER*. The principles behind the Eiffel conversion mechanism are the following:

- A conversion is an abbreviation for an object creation. You create, for example, a real from an integer, using specific rules. (For basic types there is of course no allocation of a new object, but conceptually the operation is still considered a creation.)
- Such a creation, like any other, uses a creation procedure (3.4). The class will mark some creation procedures as being also "conversion procedures".
- There is also a need for "creation functions". The reason is that instead of converting "from" we may also want to convert "to" an existing type. For example, we might want to convert a file path name, an instance of one of our classes, *PATH_NAME*, into a string. *STRING*, however, is a predefined library class; we cannot add a creation (and conversion) procedure to it. Instead we define in *PATH_NAME* a function that returns a string from an instance of this class, and mark it as a conversion function.
- We say that a type *T converts to* a type *U* if there is a conversion procedure from *T* in *U* or a conversion function to *U* in *T*.

- If an assignment *target* := *source*, or the corresponding argument passing, would not be valid according to the other rules of the language, but the type of *source* converts to the type of *target*, then — in the spirit of other mechanisms providing special syntax for important special cases — we interpret this assignment-like operation as a creation: an abbreviation for **create** *target*.*conv_proc* (*source*), where *conv_proc* is the applicable conversion procedure (or *target* := *source*.*conf_func* in the function case).
- Strict rules guarantee that there is no ambiguity. A type can convert to another in at most one way (either one conversion procedure or one conversion function); and the *conversion principle* states that a type *may not both conform and convert* to another. "Conformance" is the usual inheritance-based type relation. The principle implies for example that if *RECTANGLE* inherits from *POLYGON*, and hence conforms to it, then there cannot be a conversion procedure or function from *RECTANGLE* to *POLYGON*. (The other way around would be OK, although meaningless in this case.) As a result, no assignment will ever be ambiguous for either the compiler or a programmer.

This solution again accommodates an important classical notion within the general design rules of the language and methodology.

## 7.5 Unfolded forms

The mechanisms reviewed above enable *programs* to use different syntax for mechanisms that fit in the methodological framework. This technique has a counterpart in the definition of the *language* itself [15]: unfolded forms. The idea is that the language standard:

- For the basic forms, for example the vanilla OO feature call *x*.*f* (*args*), provides full-fledged validity rules and semantic definitions.
- For the syntactic variants, such as an operator expression *a* + *b*, defines an "unfolded form" in terms of the basic form. For example an expression of any syntax, such as the operator expression *a* + *b*, unfolds into the basic form, in this case as *a*.*plus* (*b*). Similarly, **across** … **all** … loop as well as the ∀ and ∃ variants /P17/ unfold into the basic **from** form /P15/. The validity and semantics are those of the unfolded forms.

This language description technique keeps the definition simple, since it needs only specify the validity and semantics of a core set of constructs; others follow by unfolding.

## 7.6 Agents and functional-style programming

An example of where unfolding helps is the provision of functional-programming-like facilities, embedded in the OO framework, through the concept of agent, introduced in Eiffel in 1999 [12] and a precursor to others such as delegates in C# and lambdas in Java.

We first review the non-unfolded form. Consider a routine which (conceptually) takes another routine as an argument. For example, in an interactive system using the "command-pattern" (introduced in [4]) to provide an undo-redo mechanism, in a modern version avoiding an explosion of small classes, there can be a routine

```
do_undo (doer, undoer: …)                                      -- /P20/
```

whose arguments represent the operations to execute and cancel (respectively) a certain user command. (Their types are left blank in /P20/ and will be specified below.) If we knew the command and hence the associated doer and undoer, the body of *do_undo* could just call them directly, as in

```
insert_line (editor)                                  -- /P21/
cancel_line_insertion (editor)
```

but we do not know them: *doer* and *undoer* are arguments representing the corresponding do and undo operations for an arbitrary command (that is the whole idea of the pattern). Even so it would be desirable to use the same syntax:

```
doer (editor)                                         -- /P22/
undoer (editor)
```

where *doer* and *undoer* are, as noted, formal arguments of the enclosing routine /P20/.

We note in passing that C comes close to allowing such a notation, if you are willing to add various symbols ("&", "*", parentheses). In C the arguments represent addresses for the corresponding code; you may call *do_undo* with actual arguments such as

```
&insert_line, &cancel_line_insertion                  -- /P23/
```

then *do_undo* itself can call the associated routine as

```
(*doer)(editor)                                       -- /P24/
```

The arguments in this low-level mechanism are just addresses and it is easy to circumvent type restrictions and create run-time havoc (since the given memory addresses could contain unwanted code or no code at all). Higher-level mechanisms such as Eiffel's agents and C#'s delegates provide type-safe versions of this facility.

How do we allow a routine (such as *do_undo*) to receive routines as arguments? In an object-oriented context, the only values programs ever manipulate are objects or references to objects. A routine — a piece of computation — is *not* an object; in fact the object-computation duality lies at the basis of the object-oriented method (which defines the architecture of software systems by attaching computations to object types). The solution then is to define objects that *represent* computation, and hence can be put into data structures. Such objects are called agents.

The common ancestor type of all agents is

```
ROUTINE [ARGS –> TUPLE]            -- " –>" can be read as "constrained by".
```

where *ROUTINE* is a Kernel Library class devised for this purpose. *ARGS*, the generic parameter, represents the types of arguments of an associated routine. For example instances of *ROUTINE [TUPLE [INTEGER, INTEGER]]* are agents associated with routines of two integer arguments. To understand this class specification note that:

- A generic class can have a constrained generic parameter: if its declaration starts with **class** *C [G –> CT]*, where *CT* is a type, then a generic derivation *C [T]* is only valid if its actual generic parameter *T*, also a type, conforms to *CT* (its base class is a descendant of the base class of *CT*). For *ROUTINE* the constraining type is *TUPLE*.

- *TUPLE* is a language-defined type describing sequences (tuples) of values. There are an infinity of tuple types: *TUPLE* covers all tuples; *TUPLE* [*T*], for a type *T*, covers tuples of at least one element, the first of which is of a type conforming to *T*; *TUPLE* [*T*, *U*], tuples of at least two elements starting with values conforming to *T* and *U*; and so on. So the conformance relation is: all tuple types including *TUPLE* [*T*] conform to *TUPLE*; then *TUPLE* [*T*, *U*] conforms to *TUPLE* [*T*]; and so on.

*ROUTINE* is generally used not directly but through its descendant classes

```
PROCEDURE [ARGS –> TUPLE]
FUNCTION [ARGS –> TUPLE, RES]
```

representing procedures and functions respectively. Values of those types are obtained through the **agent** notation; for example with *p* a procedure with two integer arguments and *f* a function with an integer argument and a boolean result we may use

```
agent p        -- of type PROCEDURE [TUPLE [INTEGER, INTEGER]]     -- /P25/
agent f        -- of type FUNCTION [TUPLE [INTEGER], BOOLEAN]
```

(where the types can be expressed without the *TUPLE* mentions, see /P33/ below). The basic activity of OO design and programming— look for data abstractions — again pays off: **agent** *p* is not a spiffed-up costume for a *&p* function address but truly an object in the OO sense, representing a routine equipped with applicable operations; one of these operations, seen next, is to call the routine, but it is not the only one. *ROUTINE* and its descendants have other features describing properties of the associated routines, and classes covering sequential data structures contain iterator features such as

```
do_if (action: PROCEDURE [TUPLE [G]];                              -- /P26/
        test: FUNCTION [TUPLE [G], BOOLEAN])     -- For simpler types see /P32/
    -- Apply action on all elements of the structure satisfying test.
```

Details of the agent and tuple mechanisms, and in particular the associated type system properties, are not essential for the rest of this presentation, but let us focus again on notational simplifications. For an agent *a*, the basic features include *call*, which calls the associated routine, and, for a function, *last_result* (which returns the last result produced by a call) and *item* (which calls call and returns the value of *last_result*). So the body of

```
r (a: PROCEDURE [TUPLE [INTEGER, INTEGER]];
   b: FUNCTION [TUPLE [INTEGER], BOOLEAN])
```

can call *a*.*call* ([0, 1]) and *b*.*item* ([0]) (the latter returning a boolean). Similarly, with the *do_undo* routine /P20/, which we can now specify with an appropriate signature

```
do_undo (doer, undoer: PROCEDURE [TUPLE [EDITOR]])                 -- /P27/
```

(with a syntax also soon to be simplified, see /P32/ next), a typical call will be

*do_undo* (**agent** *insert_line*, **agent** *cancel_insert_line*) -- /P28/

The body of *do_undo* may itself contain calls such as

*doer*.*call* ([*editor*]) -- /P29/

which when executed as a consequence of the call /P28/ will mean

(**agent** *insert_line*).*call* ([*editor*]) -- /P30/

and in this case has the same effect as the direct call

*insert_line* (*editor*) -- /P31/

although of course we do not know that in the body of *do_undo*, where *doer* could represent any other routine with the appropriate signature.

The notations for using the mechanism (/P25/, /P26/, /P27/, /P29/) are consistent and the underlying structures clear, but the use of *TUPLE* and brackets can make the notation tedious when used extensively. It forces the programmer to remember all the time the underlying intellectual construction, even when the goal is simply to pass a routine *p* to a routine *r* and let *r* call *p*. Once again, syntactical simplifications intervene. The first is the permission to omit the *TUPLE* specification and its brackets in /P27/, giving simply

*do_undo* (*doer*, *undoer*: *PROCEDURE* [*EDITOR*]) -- /P32/

Similarly, the types in /P25/ can be expressed more concisely as

**agent** *p* -- of type *PROCEDURE* [*INTEGER*, *INTEGER*] -- /P33/
**agent** *f* -- of type *FUNCTION* [*INTEGER*, *BOOLEAN*] -- /P34/

Since *PROCEDURE* expects a tuple type as generic parameter, the syntactical simplification does not introduce any ambiguity; /P32/ is simply an abbreviation for /P27/. The original remains usable for someone preferring a purist form. /P32/ carries the right impression: that *do_undo* expects two arguments, each of which represents a procedure with an argument of type *EDITOR*. The role of tuples is essential to make the mechanism type-safe, and consistent with the OO type system, but there is no need to force the programmer to remember every time that arguments are conceptually grouped into a tuple. The situation is similar for *FUNCTION*, which expects two parameters as in /P34/.

On the caller's side, we note in the same spirit that *call* always expects a tuple argument, enabling us to drop the brackets in *doer*.*call* ([*editor*]) (/P29/) and get

*doer*.*call* (*editor*) -- /P35/

The motivation is the same: for the programmer, there are no tuples, only arguments, in this case just one, *editor*, with no need to remember that it is internally treated as a one-element tuple. (Precautions are taken to ensure that there is no ambiguity in the case of a routine with an argument that is actually of a tuple type. Here we reap the advantages of a strong type system which enables the analysis to remove any confusion.)

The presence of "**.** *call*" is still in some sense noise since it forces the programmer to remember that the argument *doer* is of an agent type with its specific properties. In simple cases the programmer simply wants to think of *doer* as a routine (rather than the correct and more general but also more elaborate notion of an object encapsulating a routine). The type system helps again: we allow, with the guarantee of non-ambiguity enforced by type analysis, dropping the "**.***call*" part to allow, as a synonym for /P35/ and hence another synonym for the fully explicit version /P29/, the shortest possible form

*doer* (*editor*) -- /P36/

which is what we wanted in the first place (/P22/, similar to a direct call to a known routine /P31/, and to a C routine call (/P23/, /P24/) but with full type safety, a higher level of reasoning, and no need for messy pointer manipulations through & and * symbols.

This example again illustrates the general idea of devising mechanisms that take full advantage of OO concepts, particularly reliance on data abstraction, while providing familiar and concise syntactical forms.

## 7.7 About the role of agents: functional mechanisms in an OO shell

Agents call for a few more comments, not related to notation.

Functional languages have an enthusiastic following, due to their closeness to mathematical concepts and their consequent rejection of imperative constructs. It is important to note, however, that functional programming as typically practiced adds some imperative aspects, if only to model stateful operations of the program's domain (printing is an elementary but typical example). To unabashed imperative programmers, the benefit of rejecting assignment — a notion that after all has clear axiomatic models — in the name of simplicity and elegance, only later to bring in such notions as monads, possibly not the most intuitive possible for non-computer-science-PhDs, sounds dubious.

In understanding how OO and functional ideas can coexist and support each other, it is important to understand where their strengths lie. Object-oriented concepts are primarily a system structuring mechanism, based on the identification and classification of data abstractions as the backbone of a software architecture. They are unrivaled when it comes to building large systems which will be maintainable and extendible, with reusable parts. Functional programming has no particular contribution here. [16] discusses how to embed functional elements in an OO architecture. Agents provide these elements.

Since agents define abstractions of operations, including functions, much of what one can do in a functional programming language — define and manipulate functions, including higher-level functionals — is readily doable in Eiffel. So you get the best of both worlds. Not *all* of the second world; there is no direct equivalent, in particular, of the mechanism of modern functional languages, from Miranda on, for defining functions by inductive pattern matching rather than programming-style recursion. ("Here is the result if the argument is an empty list, and there it is in terms of *a* and *L* if it is the concatenation of an element *a* and a list *L*.")

# 8 EXCEPTION HANDLING

Modern imperative programming languages all have an exception mechanism, typically based on the try A catch B scheme of C++, which means: execute A, but if an exception occurs handle it in B.

## 8.1 Modes of exception use

Try-catch mechanisms fulfill the role of exceptions, if we define that role as handling abnormal run-time events. Such a definition, however, is so vague as to be almost useless:

- What is an "*abnormal event*"? What is normal to me can be abnormal to you.
- What does it mean to "*handle*" an exception? Correct the problem? Proceed as if it had not happened?

The observation of both programming textbooks and programming practice in try-catch languages shows a wildly diverse range of uses of exceptions:

- At one end, exceptions serve to recover from truly unpredictable run-time situations, such as arithmetic overflow (hard to predict in most programming languages), running out of memory, or a file suddenly becoming unavailable while being read (say someone turns off the device).
- At the other extreme, exceptions are just a control structure, another way of providing a goto while pretending (section 2) that the language supports "structured programming". An example — not made up! — is a programming style in which the routine searching for an element in a hash table produces an exception when it does not find it.

While the last example is an aberration, it is not always so clear to a practitioner when exceptions are a legitimate solution in preference to more sober control structures. Programmers need clear methodological guidelines defining the role of exceptions in the construction of programs.

Java makes a step in this direction by requiring the author of a routine (method) to specify which types of exception it may trigger and not handle by itself; then any calling routine must handle the corresponding exception types (by naming them in a catch clause). Unfortunately this obligation only applies to programmer-raised exceptions — so-called "checked" exceptions, those actually used as substitutes for goto instructions or multiple returns — and do not help for the crucial cases, which involve exceptions occurring unpredictably at run time.

## 8.2 A clear conceptual framework for exception handling

Eiffel introduced a methodology for using exceptions properly. (The first presentation in 1988 could not find a publication outlet and remained a technical report [5]; the ideas were explained in chapter 12 of [11].) The reasoning is the following.

Any operation is characterized by a contract: it *assumes* some initial properties (precondition) and *promises*, if they initially hold, to deliver some final properties (postcon-dition). The contract is the combination of these properties. That characterization applies to operations at any level, from the topmost — say a highly complex operation such as launching a rocket — down to a single routine and even further to a basic operation such as adding two integers (with the implicit precondition that the sum fits in the computer representation, and postcondition that the result expresses the mathematical sum in that representation).

The need for exception handling arises when the operation is — for any reason — not able to meet its promise. The first step towards a methodology is to distinguish between two complementary concepts:

- An **exception** is a run-time event that prevents an operation from executing the scenario — specifically, in software, the algorithm — in which it is engaged to fulfill its contract (to hold its promise).
- A **failure** is an operation's inability to fulfill its contract.

The reason the two concepts are different is that there may be more than one way to fulfill the contract. The operation may have, metaphorically, lost a battle but not the war. Exception mechanisms exist because the designer of the operation may have provided one or more *alternative scenarios* to achieve the contract in case the original one fails.

The next methodological step is to examine what courses of action make sense in the system element — in software, a routine — whose execution encounters an exception. In light of the previous observations, only two responses make sense:

- **Resumption**: start an another scenario (another way to try to ensure the contract).
- **Failure**: Give up and concede failure (accepting the loss of not only the specific "battle" but the routine's "war").

(A third strategy, "**oblivion**", is in principle possible: ignore the exception and return as if nothing had happened. It is only justified in special cases; think of a spurious exception generated by the operating system, for example after window resizing, which does not matter for a particular program. We will see below how to specify this behavior as a special case of resumption.)

The third and final methodological rule applies to the second case above: *the failure of a routine causes an exception in its caller*. (If there is no caller, for the top-level element of the execution, the execution as a whole fails.)

Any exception has a type, such as "arithmetic overflow". Up the call chain, the type of the secondary exception resulting from failed calls is "routine failure".

## 8.3 Eiffel exception mechanism

The language mechanism directly follows from the analysis. It consists of two parts:

- An effective routine can include, after its body (**do** clause), a **rescue** clause, whose instructions take over when an execution of its body triggers an exception. Termination of the rescue clause causes either failure or resumption.

- To decide between those two possibilities, an integer local variable **Retry** (predefined, like **Result** for functions from section 2) is available. If its value is 0, as it is by default, the **rescue** clause ends in the routine failing (and, as explained above, triggering an exception of type "routine failure" in its caller). Otherwise, the routine executes its body again, without reinitializing the local variables (including **Result** and of course **Retry** itself).

Here is a typical example of use:

```
transmit_if_possible (f: FILE; max: NATURAL)                          -- /P37/
        -- Attempt to transmit f over an unreliable line, in at most max attempts;
        -- produce a message if impossible.
    do
        if Retry < max then
            attempt_transmission (f )
                -- We assume that attempt_transmission, a low-level transmission
                -- routine, may cause an exception.
        else
            print_message
        end
    rescue
        Retry := Retry + 1
    end
```

This routine never fails since it can always satisfy its contract, which states that it should either transmit the file or print a message. In contrast, the version with the following shorter body (same header) can fail:

```
    do                                                                 -- /P38/
            attempt_transmission (f )
    rescue
        if Retry < max then
            Retry := Retry + 1
        else
            Retry := 0
        end
    end
```

If **Retry** reaches the maximum, the **rescue** clause resets it to zero, hence terminating with a zero value for it and causing the routine to fail. Finer control is available:

- The program has access to the exception type, enabling it to treat various cases differently. In fact, it has access to *two* exception types: the original one (such as arithmetic overflow) and the resulting "routine failure" higher up in the call chain.
- The program can also define its own exception types and, through a library routine *raise*, trigger an exception of one of these types.

- Exceptions of certain types are *ignorable*. The programmer can specify, again through a library mechanism, that execution will actually ignore (or not) exceptions of an ignorable type, thereby allowing the "oblivion" policy (8.2) when harmless.

The methodological reasoning extends to a concurrent context, raising among others the question of what should happen to an exception raised by a routine whose caller was asynchronous (in another thread or process) and may even have ceased its execution. See [21] for the answer in the SCOOP concurrency mechanism of the next section.

Clearly, the Eiffel rescue-retry style can emulate any try-catch-style exception handling. The other way around too, although only by writing fairly complicated control structures to represent resumption. The spirit is, however, different.

Like others in this article, the exception handling constructs address an important design and implementation need in conformance with principles of modern software development, and provide a methodological framework for applying the ideas.

# 9 CONCURRENCY

Concurrent programming used to be a specialist endeavor for people writing operating systems and networking applications, but is practiced today much more widely, thanks in particular to the spread of multithreading technology. Here too it is important to maintain a consistent level of abstraction by providing mechanisms that are compatible with modern methodology principles, rather than taking programmers through a roller-coaster ride between wildly different levels of abstractions.

## 9.1 Semantic mismatches

An example of the roller coaster is the typical multithreading mechanism of many OO languages. While some modern languages (such as Go) were designed specifically for concurrency, much of the concurrent programming performed in object-oriented languages uses "thread libraries", which provide low-level facilities, relying in particular on semaphores as the principal means of synchronization.

The semantic gap is enormous. OO languages emphasize abstraction; their control structures, even with remnants of the goto, limit the potential for "spaghetti code"; if they are statically typed, the type system provides powerful means of expression and eradicates many errors. In contrast, the semaphore is a concept from the 1960s, a goto-like low-level construct opening the way for many concurrency errors. It makes it hard in particular to avoid **race conditions**: cases of data corruption, leading to incorrect results, due to different threads writing and accessing the same data, each in an order that is correct for the thread taken independently but incorrect in combination with the actions of the other threads (as in a carelessly designed reservation system that presents the same seat to two customers as available, and lets them both reserve it).

## 9.2 An example: what happens to preconditions?

Initial inquiries that led to the design of Eiffel's concurrency mechanism, SCOOP, examined [9] the following question, a specific case of the race condition issue: what does a precondition mean in a concurrent context? Consider a typical precondition-equipped routine, inserting an element into some data structure *s*:

```
put (s: STRUCTURE; e: ELEMENT)                                    -- /P39/
        require
            not s.is_full
        do
            … Implementation of the insertion (see below)…
        ensure
            s.has (e)
            not s.is_empty
        end
```

The precondition in /P39/ entitles the code for "Implementation of the insertion" to include operations such as

```
s.insert (x)                                                      -- /P40/
```

where *insert* itself also has the precondition **not** *s.is_full*, since this operation is executed as part of *put* whose precondition in /P39/ guarantees that same property.

Under the basic ideas of Design by Contract, the precondition of a routine such as *put* is not just a *requirement on* the caller but also a *guarantee to* the caller that it is the *only* requirement: the supplier (here the routine *r*) *commits* to handling correctly any call that satisfies the precondition. Examples of such calls include

```
if not my_struct.is_full then put (my_struct, x) end              -- /P41/
```

and

```
remove (my_struct) ; put (my_struct, x)                           -- /P42/
```

assuming that *remove* (dually with *put*) has the postcondition **not** *is_full*. Specifically, the code marked "… Implementation of the insertion …" can safely rely on the property **not** *s.is_full*, part of the precondition and hence guaranteed by the caller.

Such modes of reasoning are fundamental to our ability to write meaningful programs: if a supplier states a condition for doing its work, and you can satisfy that condition, then you will get the promised result.

In a concurrent settings, this pillar of civilization crumbles!

If the client (the code calling *put* in /P41/ or /P42/) and the supplier (the code for the routine, executing such operations on the structure as *insert* in /P40/) are running in different threads but have access to the same structure (here under different names, *my_struct* and *s*), the guarantee evaporates. Between the time you ascertain **not** *my_struct.is_full* in /P41/ (or complete the call to *remove* in /P42/) and the time you call *s.put* (*x*), some other thread may access *s* and make it full again.

The solution to this "precondition paradox" in SCOOP appears in section 9.8 below. More generally, this discussion illustrates the spirit that presided over the design of a concurrency mechanism for Eiffel, in line with the general goals.

## 9.3 Sequential versus concurrent

Some approaches to concurrency start from the position that everything should be concurrent by default; sequential computing is just a special case. While this argument is by itself correct, it does not help much in practice. Concurrent reasoning is difficult. The development of programming methodology over the past decades has reached a stage at which we master the processes of sequential programming to a good extent; bringing unrestrained concurrency into the picture makes things far more difficult.

The precondition example is a typical case. We are used to thinking that if it has been ascertained that operation *B* will perform correctly under condition *C*, and that *A* ensures *C*, we may comfortably execute *A* then *B*, as in /P42/. With the interleaving of operations permitted by concurrency, in particular by multithreading, such inferences are no longer justified. No other form of reasoning presents itself as an obvious replacement.

As a consequence of these observations SCOOP lets the programmer, as much as possible, *continue to use the modes of reasoning that work in sequential computing*, circumscribing concurrent aspects to specific departures from sequential mechanisms. In addition, SCOOP provides expressive power not usually available in other frameworks.

## 9.4 Simultaneous resource reservation

An important example of added expressive power is the ability to reserve several resources at once. The implementation of the classic dining-philosophers problem reads, in the scheme for each philosopher:

```
separate left, right as l, r do                                    -- /P43/
        l.pick ; r.pick ; eat (l, r) ; l.put_down ; r.put_down
end
```

The **separate** construct gets exclusive hold of the objects listed, giving them local names (*l* and *r*). The body of the construct can then manipulate these objects safely — as in a sequential context — since they are under its exclusive control. (For completeness of the example we may assume that *eat* has preconditions *l.picked* and *r.picked*, *pick* has a postcondition *picked*, and *put_down* a postcondition **not** *picked*.)

/P43/ (in a loop for each philosopher) is essentially the entire solution to this classic problem, shorter and simpler than those traditionally proposed. The ability to obtain exclusive hold of an arbitrary set of objects frees the programmer from complicated, error-prone (in particular, deadlock-prone) algorithms; the mechanism is handled by the SCOOP runtime, providing programmers with a significant gain in expressive power.

## 9.5 Processors and regions

"Resource" as used above is not a special language concept. Neither do we need any special notion of "concurrent object". SCOOP simply uses Eiffel objects; the only fundamental new concept for concurrency is that of a "processor".

Introducing active objects, as in some OO concurrency frameworks, violates the basic scheme of object-oriented programming, in which an object is a server, providing a number of services (features) to other objects, its clients, through feature calls. Such an object cannot have its own scenario (as an active object would have), since it would immediately clash with the scenarios of its clients when they call its features; these clashes lead to complicated synchronization mechanisms.

Concurrent OO programming does not need any of that. It suffices to extend the sequential framework by adding that each object has an associated **processor**, a sequential mechanism in charge of executing all the features called on the object. The model is general enough to accommodate various realizations of the notion of processor; it could be a physical CPU or a software process. In current implementations, processors are threads.

A processor is sequential, like a traditional CPU; concurrency comes from having several processors. The first major difference between sequential and concurrent computation is this move from one to multiple processors.

## 9.6 Synchrony, asynchrony and wait by necessity

The fundamental OO operation calls a feature *r* on a supplier object S, on behalf of a client object C (which knows S under some name *x*, for example an attribute name):

*x*.*r* (…) -- /P44/

Since all such calls on a given S are handled by S's processor, we may see the set of objects as partitioned into a number of "regions", each associated with a processor.

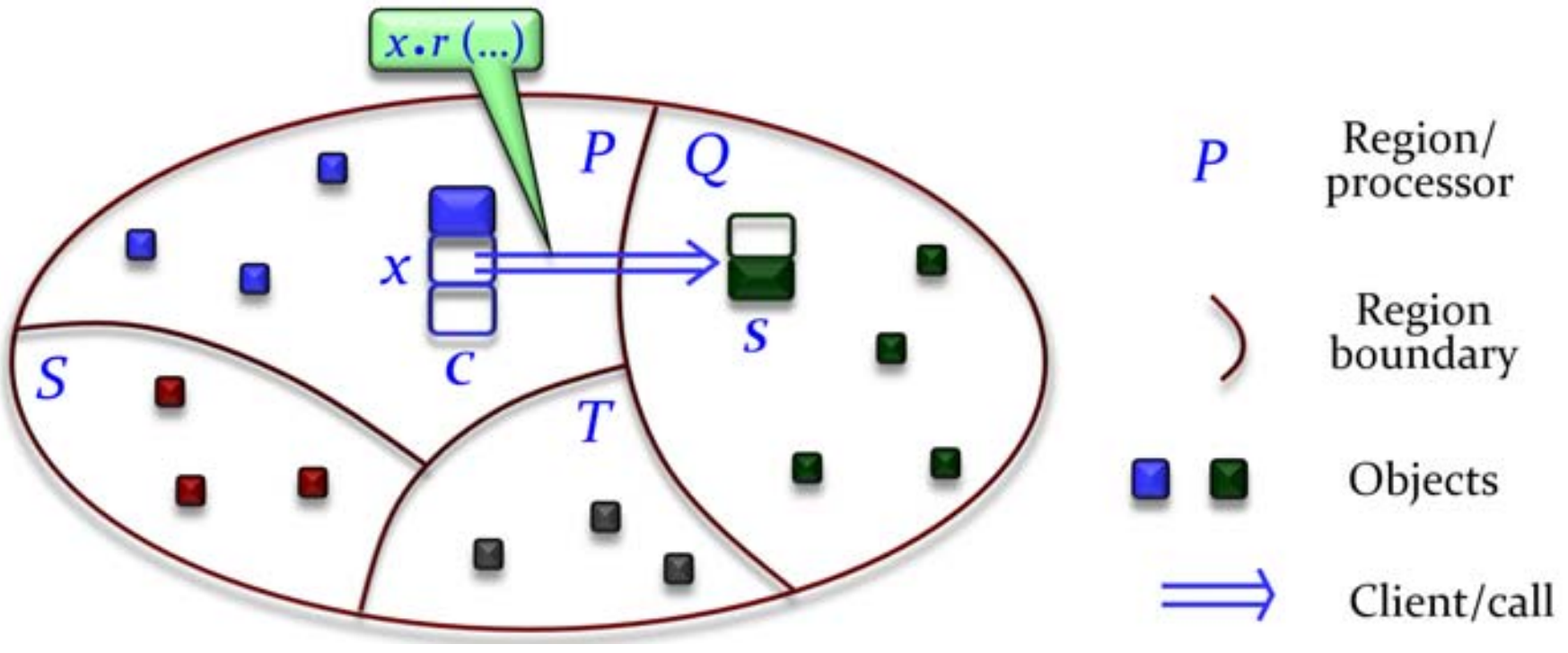

The second major difference between sequential and concurrent comes with the semantics of calls such as /P44/. The call will be started by the processor *P* handling *C*; it requests the execution of the routine *r* on the supplier object *S*. Let *Q* be the processor handling that object. We say that the call is non-separate if *Q* is *P* (the two objects are in the same region), and separate if they are different processors, as in the figure.

Considering the effect of the call:

- For a non-separate call, only one processor is available; the only possibility is the same as in sequential programming (which has just one processor available for the entire computation). The call execution has to be *synchronous*: the client does not proceed until the body of *r* has executed to completion.
- For a separate call, *asynchrony* is possible: the client's computation can continue, without waiting for the execution of *r* to complete.

The asynchronous case leads to a distinction (due to Benjamin Morandi in his ETH PhD thesis, bit.ly/3XSIpfx) between two parts of the semantics of a call such as /P44/: *call* (proper) and *application*. The caller executes the call, which means that it registers its request to have *r* executed; the supplier executes, or "applies", the body of *r*. For synchronous calls, including sequential programming, the difference does not matter, since call causes immediate application; with asynchrony, the client can proceed with further instructions immediately after having executed the call.

The relationship is subject to the following consistency rules:

- **Correctness**: at the time an application of a routine *r* ends, the postcondition of *r* holds.
- **Causality**: the application of a call follows it. (The application may occur at any time after the call, but not before.)
- **Order preservation**: if two calls involve the same client processor *and* the same supplier processor, their applications happen in the same order as the calls.

Order preservation is limited to this case of interaction between the same two processors: no order guarantee exists for the application of calls from different clients and suppliers. In other words, there is no need (which would be highly restrictive) for a global clock.

Separate calls permit asynchrony, but do not require it: a synchronous semantics would still be possible (although it might lead to deadlocks). When are calls actually asynchronous? The answer makes it possible to address another question: how does a client catch up with an asynchronous supplier? If the client started a call, such as *New_York*.*sell* ("*Microsoft*", 300), it is for a purpose (in this case, disposing of some shares); the client may need at some point to make sure that the purpose has been achieved, in other words that the application —not just the call — has executed to completion. It does so by executing a *query* (A.7, 1.1), for example *New_York*.*portfolio* ("*Microsoft*") (to find out the current number of shares). The rule, known as **wait by necessity** (it could also be called "lazy wait" and is originally due to Denis Caromel) is that waiting for resynchronization occurs in this case, and in this case only. The call will cause the client to wait until all previous separate calls with the same target (here *New_York*) have completed their application, and the application of the current query (*portfolio*) is also complete.

For a separate call:

- If the routine is a command (or procedure, see A.7), the call is asynchronous.
- If it is a query (function or attribute, see 1.3), the call is synchronous.

A typical pattern consists of firing a succession of command calls to tell suppliers to perform operations, then executing query calls to take advantage of the results. (In the example, send all your buy/sell orders, then find out where your portfolio now stands.)

## 9.7 Language support

Since separate and non-separate calls have different semantics (possibly asynchronous versus guaranteed synchronous), the source text must express which version applies. The type system takes care of this matter as of many others. A declaration of the form

```
x: separate T                                          -- /P45/
```

(as opposed to the default *x*: *T*) states that the object denoted by *x* may be separate (reside in a different region, handled by a different processor). Procedure calls of target *x* may then be asynchronous (their application is decoupled from the call), and query calls of target *x* will force the client to wait until the query completes.

A **separate** declaration such as /P45/ does not specify the processor of *x* and does not even require it to be a different processor: separate means "*potentially* separate". The processor could actually be the same in some executions. What the declaration specifies is that asynchronous execution is *possible* (depending on processor circumstances) for calls of target *x*.

The extension of Eiffel from a sequential programming language to a concurrent one uses only one keyword, **separate** (also used in the resource reservation construct /P43/, although in that case, as seen next, only as an abbreviation).

In line with the general observation that the properties of the language are largely embodied in the type system (A.5), type rules are associated with the preceding concepts: in particular, you can assign from non-separate to separate (remember that separate means *possibly* in a different region) but not the other way around. Associated type rules apply to calls and were elegantly defined in Piotr Nienaltowski's ETH PhD thesis (se.ethz.ch/people/nienaltowski/papers/thesis.pdf).

## 9.8 Synchronization

Consider a routine whose formal arguments include some having separate types, as in an adaptation of /P39/ using a separate argument to denote a shared structure:

```
put (s: separate STRUCTURE; e: ELEMENT)                -- /P46/
        require
            not s.is_full
    … The rest as in /P39/ …
```

Then a call such as *put* (*my_struct*, *x*) will proceed when both:

S1 The processors corresponding to all separate arguments, in this case just *my_struct*, are available.

S2 These arguments satisfy the corresponding *separate preconditions*.

A “separate precondition” is a precondition clause containing an expression *x.some_property* where *x* is a separate formal argument, as is the case with the precondition here.

This rule provides the basic synchronization tool. It is in line with the discussion of section 9.2 and answers the question raised there, “what happens to preconditions?”. Non-separate preconditions retain their sequential semantics of *correctness* conditions, but separate preconditions such as the one for *put* in /P46/ become *wait* conditions. It is important to point out that this decision is not arbitrary; it appears to be the only reasonable choice, since the standard correctness-condition semantics is just impossible to apply, as shown in the discussion of 9.2. It also provides a powerful synchronization mechanism; note in particular that the scheme for *put* in /P46/:

```
put (s: separate STRUCTURE; e: ELEMENT)                    -- /P47/
        -- Insert e into s.
    require
        not s.is_empty
    do
        … Implementation of the insertion …
    ensure
        not s.is_empty
    end
```

provides (together with the dual scheme for *remove*) the basic implementation of a producer-consumer system with shared buffer, expressed very simply.

An important component of this semantic specification is that the reservation of resources corresponding to rule S1 above (wait until all processors for separate arguments are available) applies to *any number* of separate arguments. A philosopher object may for example call *step (left, right)* with the routine definition

```
step (l, r: FORK)                                          -- /P48/
        -- Execute the basic cycle step of a philosopher with l and r.
    do l.pick ; r.pick ; eat (l, r) ; l.put_down ; r.put_down end
```

providing a variant of the implementation using the **separate** construct /P43/. After a call to *step*, application of the routine’s body will only proceed when it gets exclusive hold on *both* processors for the arguments. As was noted in the presentation of the **separate** construct, this facility achieves high expressive power by providing a mechanism for the simultaneous reservation of multiple resources. The soundness of the mechanism is guaranteed by the rule which actually *requires* a separate target *x*, in a call *x.r (…)*, to be a separate formal argument of the enclosing routine, or a local name defined in an enclosing **separate** construct.

The form with the feature call, /P48/, is actually the basic construct. The **separate** construct (in the /P43/ example, **separate** *left*, *right* **as** *l*, *r* **do** *eat* (*l*, *r*) **end**, is just an abbreviation for the call *step (left, right)* and the routine definition of *step* in /P48/, in the spirit of "unfolded forms" (7.5). It makes it possible, when an operation needs to work on separate expressions, to avoid writing a wrapper routine, such as *step*, for it. But the semantics is the same. Wrapping the processing in a routine remains necessary in the case of wait conditions, which the routine will express as separate preconditions.

SCOOP has turned out to be a practical and safe solution to concurrent programming using threads, letting programmers apply the same principles as in the rest of OO design and programming with Eiffel.

## 10 VOID SAFETY

The description of the last mechanism in this review will be only a sketch, even though it is one of the most important characteristics of today's Eiffel.

Eiffel is void-safe: a null-pointer dereference cannot occur.

The dereferencing of a null pointer (C terminology), or of a void reference (OO terminology, for a more abstract notion of pointer) is an attempt by the execution of a program to execute or evaluate *x*.*a* for an attribute *a* or *x*.*r* (...) for a routine *r* in a state in which *x*, a pointer or reference is null, or void, meaning that it does not denote any object. Null-pointer dereferencing is also known as a "void call".

In C, in the absence of a full-fledged exception mechanism, the consequence of a void call is to crash the program. In exception-equipped languages such as C++, Java or C# (or Eiffel prior to the introduction of void safety) a void call causes an exception, which often has the same result anyway in the absence of a handler specifically written for that case. (The reason programs typically do *not* catch such exceptions is that if the programmer realizes that *x* can be void, it is much easier to fix that problem, handling the case of a void *x* through an ordinary test in a conditional instruction, than to let the exception happen and try to recover from it. As a result, the void calls that do remain are the tricky ones that the programmers did not foresee, directly leading to a crash. Being tricky, they are also harder to debug.)

In the Common Vulnerabilities and Exposure (CVE) database, hundreds of the most egregious cases of security violations, some in widely used products from such companies as Apple, Cisco, Google, IBM and Microsoft and technologies such as JPEG involve null-pointer dereferencing, as analyzed in Alexander Kogtenkov's ETH PhD thesis (se.ethz.ch/people/kogtenkov/thesis.pdf, which also proved the soundness of the Eiffel scheme). CVE mostly refers to C programs, but the problem persists in all the dominant OO languages, including the most strictly statically-typed ones. Some recent designs, such as Kotlin, provide a partial solution, but we know of no other widely available OO language than Eiffel that provides a full solution, mechanically proven sound.

One should note in passing that functional programming does not solve the problem. Functional languages do address it in part by removing the notion of pointer or reference and providing instead built-in classes for common data structures such as lists, hiding pointer manipulations from the programmer. But efficient modeling of complex data structures not covered by the library may require a pointer-like mechanism.

The Eiffel solution (which was initially influenced by early attempts in research languages, notably C-omega from Microsoft Research) was initially described in [13], with the ironed-out and fully-implemented version explained a few years later in [19] and Kogtenkov's thesis. Like many other critical mechanisms (see the comment at the end of A.5), it essentially consists of enriching and tightening the type system.

The starting point was the realization that for reference types void (or null) is an anomaly, not the more common case. The reason programming languages typically accept that references can be void is computer scientists' familiarity with linked data structures, for which it is customary (and often appropriate) to terminate a structure by linking the last element to void. In application domains, however, there is usually no real need or room for void values. If we have a class *EMPLOYEE* in a payroll system, what is the point of a void *EMPLOYEE* value? None in fact.

As a result of this observation, we consider that reference values should by default be "attached", meaning guaranteed always to denote an object at run time; the case of a "detachable" expression, which can be void, remains the exception. So if we declare

*e*: *EMPLOYEE* -- /P49/

we are making a *commitment* that the corresponding object or objects will always exist. If we want to allow an expression to become void, for example for a list cell, we say it explicitly by declaring it as

*next*: **detachable** *CELL* -- /P50/

(In /P49/ we may use the explicit form **attached** *EMPLOYEE*, but this is not necessary since "attached" is the default for the reasons noted.)

In the attached case, /P48/, it is our responsibility to meet the commitment, and of course the compiler's responsibility (backed by the language definition and the aforementioned formal proof of soundness) to enforce it, guaranteeing that *e* will never be void when execution attempts to apply a feature to it. The precise rules will not be described here (see the cited references and the language standard), but it is important to mention the two principal properties:

- As noted, the type system is key. The basic rule (compare with the SCOOP rule in 9.7) is that you can assign from attached to detachable but not the other way around.
- The most delicate part has to do with initialization. Every object should be initialized upon creation; for fields not explicitly initialized by the creation procedure, the language relies on default values for each type (section 2), including **Void** for detachable types, but for an attached type other than basic types there is no universal default; the program must provide it.

The crux of the mechanism's viability is its convenience. An extreme solution, forcing programmers to protect every call *x*.*r* by a test for the possibly vacuity of *x*, might be theoretically suitable. It is, however, of the "straitjacket" kind and not palatable in practice. (Also, it only pushes the problem, as you have to write the "**else**" part of the condition instruction.) On the other hand, we cannot just rely on the assumption that the compiler is "smart" and finds out which cases are harmless (preclude void calls): the language is defined by a precise specification, not the behavior of any particular compiler. The rules have to be simple, and equally clear to the programmer and the compiler writer.

This need for a delicate balance between expressiveness and safety, constrained by the requirement to have a crystal-clear language specification and a provably sound mechanism, explains the delay of almost a decade between the time the essential components of the solution were first published ([13] from 2005) and when the mechanism became a fully integrated part of the Eiffel approach. The challenge was to guarantee void safety without imposing an undue burden on the programmer.

The discussion was biased initially by the difficulty of *converting* existing non-void-safe code to void safety. Once the key elements of existing software, in particular many thousands of lines of library code, were made void-safe, the focus could turn to the more important and, as it turns out, more easily attainable goal: writing code that is void-safe in the first place. Once one has got the knack of void-safe programming, this task becomes natural. While it requires a specific mode of thinking, with a new tack on OO programming, it is not hard to practice and leads to programs that are more readable and convincing — in addition to providing a major guarantee against a risk that existed in pre-void-safe Eiffel and still does in other approaches: the risk that a program that seems to have been carefully designed, seriously reviewed and extensively tested will some day, unpredictably, make a catastrophic attempt to dereference a null pointer.

## APPENDIX A1: LANGUAGE DESCRIPTION

Separate from the design of individual language constructs, Eiffel distinguishes itself by the design of the language's own definition. The Ecma and ISO standard [15] follows a strict discipline of differentiating between four levels (previewed in A.7):

- The lexical level: individual tokens.
- The syntactic level: program structure (for a program made of correct tokens). The syntax description uses BNF-E, a variant of BNF restricting, for simplicity and readability, the definition of each construct through a single production which is one of: choice, concatenation or iteration. (Usual BNF can mix these variants, for example by defining a construct as a choice between concatenations, as in Z = (U V) | (X Y), which in BNF-E requires three productions, one defining Z as A | B, one A as U V and one B as X Y.)

- The static semantic level: validity constraints (for a program that is syntactically correct). A validity constraint is a static requirement on program texts that is not expressible (or not conveniently expressible) in BNF. Type rules are a typical example.
- The semantic level: effect of constructs (if they are valid).

The figure below, an extract from the standard, shows a typical use of the last three levels. Two further kinds of elements appear in the standard:

- Informative text: explanations, often including programming examples. Although not formally part of the language definition, they help make the standard readable.
- Definitions (not appearing in the example), introducing auxiliary notions — at any of the four levels — in a precise way.

**8.9.7 Syntax: "Old" postcondition expressions**

Old $\triangleq$ **old** Expression

**8.9.8 Validity: Old Expression rule** **Validity code:** *VAOX*

An Old expression *oe* of the form **old** *e* is valid if and only if it satisfies the following conditions:

1. It appears in a Postcondition part *post* of a feature.
2. It does not involve **Result**.
3. Replacing *oe* by *e* in *post* yields a valid Postcondition.

*Informative text*

**Result** is otherwise permitted in postconditions, but condition 2 rules it out since its value is meaningless on entry to the routine. Condition 3 simply states that **old** *e* is valid in a postcondition if *e* itself is. The expression *e* may not, for example, involve any local variables (although it might include **Result** were it not for condition 2), but may refer to features of the class and formal arguments of the routine.

*End*

**8.9.9 Semantics: Old Expression Semantics, associated variable, associated exception marker**

The effect of including an Old expression *oe* in a Postcondition of an effective feature *f* is equivalent to replacing the semantics of its Feature_body by the effect of a call to a fictitious routine possessing a local variable *av*, called the **associated variable** of *oe*, and semantics defined by the following succession of steps:

1. Evaluate *oe*.
2. If this evaluation triggers an exception, record this event in an **associated exception marker** for *oe*.
3. Otherwise, assign the value of *oe* to *av*.
4. Proceed with the original semantics.

This specification discipline particularly applies to the validity constraints (static semantics), where it prescribes an "*if and only if*" style as illustrated in the extract shown for the "Old expression" construct above. All programming languages (even those without a particularly strong type system) have validity rules, stating that certain syntactically correct permutations are permitted and others not. It is easy and tempting for the language designer to focus on the "*only if*" part, by stating certain obligations that the program must meet. For example, the type of *e* in an assignment *v* := *e* *must* conform to the type of *v*. It is also natural to give *examples* that are valid and others that are not. But a good language specification should aim higher: rules should specify precisely — for the benefit of both programmers and compiler writers — **when exactly** a tentative program element is valid, and when not. Hence the "*if and only if*" style, illustrated by the VAOX rule above which states under what exhaustive conditions an "Old expression" is valid.

While it is easy to tell programmers pieces of what they may do and especially may not do (use an Old expression outside of a postcondition, use **Result** in it), the "*if*" part is harder since it represents a commitment by the language definition, which the compiler writer must honor: *if* you meet all my conditions, *then* I promise, as part of my (the compiler writer's) contract with you (the programmer) that I will be able to process your program properly so that its execution will behave as promised in the "Semantics" part.

While tough to apply, this discipline is beneficial not only for programmers but also for the quality of the language definition, which it encourages to leave no stone unturned.

Language definitions for today's principal languages do not typically follow such a discipline. The usual approach is to list some conditions, in an "*only if*" style, without a guarantee that they are *all* the conditions. In general, such standards are not particularly rigorous. The Eiffel specification is not formal in the sense of a mathematical specification, but through the traits discussed here it is influenced by practice with formal specifications, sharing their focus on precision and retaining some of their benefits.

## APPENDIX A2: LANGUAGE EVOLUTION

The final observation on Eiffel's conceptual underpinning addresses change. Discussions of language design typically present a static view: how do we build the ideal language? In practice, no one reaches perfection, or even an earthly approximation thereof, the first time around. Unless the language is a one-off shot intended to produce papers rather than programs, it continues to evolve, as programmers complain about infelicitous features, language designers themselves realize that some of their decisions were not optimal, other languages come up with mechanisms worthy of emulation, the evolution of the IT industry brings in new possibilities and needs, and more generally as new ideas come up. Languages such as Eiffel, C++, Java and C# have all been around for several decades, inevitably undergoing significant evolution during that time.

The key question, if a language has an installed base, is how to balance innovation with compatibility. The dilemma is not specific to programming languages but plagues the IT industry as a whole; it could be phrased cynically as how not to punish your users too harshly for their crime of trusting you too early.

The larger the installed base, and the more innovative the new solutions, the harder the decisions. (This matter was discussed at some length in the preface of the original language book [6].) Often, the language team resolves this dilemma in favor of compatibility: do not harm the installed base, do not scare your users. While the reasons are understandable and even commendable, the result after years of evolution is that a language can continue to carry the accumulated sediments of many earlier eras.

It will come as no surprise that a number of the mechanisms in today's Eiffel, including several discussed in this article, were not present in the initial design. The most significant update, discussed in section 10, was to make the language void-safe. But others, if less momentous, were also drastic innovations at the time of their introduction. While maintaining compatibility has been a constant concern in the evolution of the language, another strong principle is that it is important, in the spirit of the scientific method, to admit mistakes — meaning here design decisions that in retrospect turn out not to be ideal — and correct them, rather than clinging to old ways of doing things. This approach has turned out to work, *provided* language changes apply the following principles:

- *Certainty*. Only introduce an incompatible change if you are absolutely certain that the new way is better than the old; not only better, but significantly better. (If not, "*leave good enough alone*" remains a reasonable guideline.) Otherwise, language evolution would go back and forth between contradictory decisions. Another way to state this principle is that compatibility remains the norm; incompatible change is the exception which must be fully justified.
- *Discussion*. Conduct a dialog with the user community and integrate its feedback. In our experience that feedback has sometimes led to abandoning proposed changes, but for the retained ones it has often led to crucial improvements of the proposals.
- *Education*. Explain and justify again and again.
- *Transition*. Whenever possible, provide the old mechanism along with the new, at least for a generously scheduled transition period (such as two years). In general it is preferable to remove the old scheme eventually, for fear of causing the sedimentation problem mentioned above, but one should give users enough time to adapt.
- *Conversion*. Offer tools to update existing software automatically, particularly for changes that do risk breaking existing code.

The application of these principles has allowed a bolder than usual approach to language evolution, helping programmers to move with the times while preserving the essential simplicity and consistency of the language, embodied by the design principles of which this article has presented some of the most significant applications.

*Acknowledgments*: to Jean-Pierre Jouannaud and other reviewers for important comments.

## BIBLIOGRAPHY

The following list only includes publications by the author and collaborators, since they are the ones cited in the text. Those publications themselves rely on many contributions from many authors over many decades. The bibliography of [11], for example, contains 300 references to contributions of other authors.

The references throughout the text to properties of languages such as Simula, Smalltalk, C++, Java and C# come from the standard documents and textbooks for these languages. References to Eiffel properties come from the ISO standard [15] and the language site eiffel.org. Information about the EiffelStudio IDE is available at eiffel.com.